\documentclass[useAMS,usenatbib]{mn2e_mod}
\usepackage{natbib}
\usepackage{graphicx}
\usepackage{amsmath}
\usepackage{amssymb}
\usepackage{deluxetable}
\usepackage{longtable}
\usepackage{url}
\usepackage{array}
\usepackage{graphicx,kantlipsum, rotating}

\let\ACMmaketitle=\maketitle
\renewcommand{\maketitle}{\begingroup\let\footnote=\thanks \ACMmaketitle\endgroup}

\title[multi-phase outflows in E+A galaxies]{Multi-phase outflows in post starburst E+A galaxies -- I. General sample properties and the prevalence of obscured starbursts}

\author[Baron et al.]
{Dalya Baron$^{1}$\thanks{dalyabaron@gmail.com},
Hagai Netzer$^{1}$,
Dieter Lutz$^{2}$,
J. Xavier Prochaska$^{3}$ \&
Ric I. Davies$^{2}$
\\
\\
$^{1}$School of Physics and Astronomy, Tel-Aviv University, Tel Aviv 69978, Israel.\\
$^{2}$Max-Planck-Institut für Extraterrestrische Physik, Giessenbachstrasse 1, 85748 Garching, Germany.\\
$^{3}$Department of Astronomy and Astrophysics, UCO/Lick Observatory, University of California, 1156 High Street, Santa Cruz, CA 95064, USA.
}

\begin{document}

\maketitle

\label{firstpage}
\begin{abstract}

E+A galaxies are believed to be a short phase connecting major merger ULIRGs with red and dead elliptical galaxies. Their optical spectrum suggests a massive starburst that was quenched abruptly, and their bulge-dominated morphologies with tidal tails suggest that they are merger remnants. AGN-driven winds are believed to be one of the processes responsible for the sudden quenching of star formation and for the expulsion and/or destruction of the remaining molecular gas. Little is known about AGN-driven winds in this short-lived phase. In this paper we present the first and unique sample of post starburst galaxy candidates with AGN that show indications of ionized outflows in their optical emission lines. Using IRAS-FIR observations, we study the star formation in these systems and find that many systems selected to have post starburst signatures in their optical spectrum are in fact obscured starbursts. Using SDSS spectroscopy, we study the stationary and outflowing ionized gas. We also detect neutral gas outflows in 40\% of the sources with mass outflow rates 10--100 times more massive than in the ionized phase. The mean mass outflow rate and kinetic power of the ionized outflows in our sample ($\dot{M}\sim 1\, \mathrm{M_{\odot}/yr}$, $\dot{E}\sim 10^{41}\, \mathrm{erg/sec}$) are larger than those derived for active galaxies of similar AGN luminosity and stellar mass. For the neutral outflow ($\dot{M}\sim 10\, \mathrm{M_{\odot}/yr}$, $\dot{E}\sim 10^{42}\, \mathrm{erg/sec}$), their mean is smaller than that observed in (U)LIRGs with and without AGN. 

\end{abstract}


\vspace{1cm}
\section{Introduction}\label{s:intro}

The process of active galactic nuclei (AGN) feedback is usually invoked to explain various observed properties of local massive galaxies, including their total mass, the enrichment of their surrounding circumgalactic medium (CGM), and the observed correlation between the stellar mass and the mass of the central supermassive black hole (SMBH; e.g., \citealt{silk98, fabian99, king03, springel05b, kormendy13, nelson19}). AGN feedback couples the energy that is released by the accreting SMBH with the gas in the interstellar medium (ISM) of its host galaxy. It operates via two main modes, which are related to the type of nuclear activity (e.g., \citealt{croton06, alexander12, fabian12, harrison18}). The kinetic (or jet) mode is typically associated with low power AGN. In this mode, the plasma in the radio jet is the main source of energy, preventing the gas surrounding the galaxy from cooling down. The radiative (or quasar) mode is typically associated with high luminosity AGN accreting close to their Eddington limit. In this mode, the energy that is released by the accretion disc drives gas outflows that can reach galactic scales, destroy molecular clouds, and/or escape the host galaxy (e.g., \citealt{faucher12, zubovas12, zubovas14}). Such outflows are believed to have a dramatic effect on their host galaxy, capable of quenching its star formation and transforming it from a star forming (SF) galaxy to a red and dead elliptical (e.g., \citealt{dimatteo05, springel05, springel05b, hopkins06}). A number of models have successfully reproduced the correlations between SMBHs and their hosts, by requiring AGN-driven winds to carry a significant fraction of the AGN energy (5--10\% of $\mathrm{L_{AGN}}$; \citealt{fabian99, tremaine02, dimatteo05, springel05b, kurosawa09}). 

Recent simulations and observations question the relative importance of AGN-driven winds in the context of galaxy evolution. For example, recent zoom-in simulations suggest that these outflows can escape along the path of least resistance, and thus have a limited impact on the ISM of the host galaxy (e.g., \citealt{gabor14, hartwig18, nelson19}). Observations of local quasars, many of which show jets and/or outflows, reveal no difference in the molecular gas properties of the quasar hosts compared to the general galaxy population (e.g., \citealt{jarvis20, yesuf20b, shangguan20}). This suggests that quasars driving galactic-scale outflows do not have an immediate effect on the molecular gas reservoirs in the systems. In addition, recent observations of outflows in typical active galaxies suggest that the energy that is carried out by the wind is several orders of magnitude lower than the theoretical requirement of 5--10\% of $\mathrm{L_{AGN}}$ (e.g., \citealt{bae17, baron19b, mingozzi19, santoro20, davies20, revalski21, ruschel_dutra21}; see recent reviews by \citealt{harrison18, veilleux20}). 

Results about AGN-feedback at high redshifts are still ambiguous. Earlier studies focused on the most extreme objects or outflow cases, and were limited to small sample sizes. (e.g., \citealt{nesvadba08, cano12, perna15, brusa15}). Later studies present larger samples with more typical high-redshift systems, with an increased observational sensitivity (e.g., \citealt{harrison16, forster_schreiber18, leung19, kakkad20}). However, various uncertainties involved in the estimation of the mass and energetics of these winds have not yet allowed to reach firm conclusions (see discussion by e.g., \citealt{harrison16, leung19}).

While it appears as though AGN-driven winds are not powerful enough in typical local active galaxies, they might be more significant in specific, perhaps short phases in galaxy evolution. One example is the post starburst phase. Post starburst E+A galaxies are believed to be starbursts that were quenched abruptly. Their optical spectrum shows stars with a very narrow age distribution. It is typically dominated by A-type stars, without any contribution from O- or B-type stars, suggesting no ongoing star formation \citep{dressler99, poggianti99, goto04, dressler04, french18}. The estimated SFRs during the recent bursts are high, ranging from 50 to 300 $\mathrm{M_{\odot}/yr}$ \citep{kaviraj07}, and the estimated mass fractions forming in the burst range from 10\% to 80\% of the total stellar mass \citep{liu96, norton01, yang04, kaviraj07, french18}. Some of these systems show bulge-dominated morphologies with tidal features, suggesting that they are merger remnants \citep{canalizo00, yang04, goto04, cales11}.

Different studies suggest that E+A galaxies are the evolutionary link between gas-rich major mergers (observed as ULIRGs) and quiescent ellipticals (e.g., \citealt{yang04, yang06, kaviraj07, wild09, cales11, cales13, yesuf14, wild16, baron17b, baron18, french18}). Inspired by the observed connection between ULIRGs and quasars (e.g., \citealt{sanders88}), hydrodynamical simulations of galaxy mergers suggested the following evolutionary scenario. A gas-rich major merger triggers a starburst that is highly obscured by dust and is primarily visible in far infrared (FIR) or sub-mm wavelengths. At this stage, gas is funneled to the vicinity of the SMBH, triggering an AGN. The AGN then launches powerful outflows that sweep up the gas and remove it from the system, thus quenching the starburst abruptly (e.g., \citealt{springel05, hopkins06}). 

Several observations challenge this simple picture. First, dusty starbursts can exhibit E+A-like signatures (strong H$\delta$ absorption and weak line emission; see \citealt{poggianti00}), and can thus be mistakenly classified as quenched post starburst galaxies. Concerning AGN feedback, several studies find large molecular gas reservoirs in some E+A galaxies (e.g., \citealt{rowlands15, french15, yesuf17, french18, yesuf20}), suggesting that the AGN did not remove or destroy all the molecular gas in the system. Other studies find a delay between the onset of the starburst and the peak of AGN accretion (e.g., \citealt{wild10, yesuf14}).

Although some studies presented indirect evidence for AGN feedback taking place in E+A galaxies (e.g., \citealt{kaviraj07, french18}), little is known about this process, in particular about AGN-driven winds, in this short-lived phase. \citet{tremonti07} found high-velocity ionized outflows, traced by MgII absorption, in z$\sim 0.6$ post starburst galaxies (see also \citealt{maltby19} for a more recent study at z$\sim$1). However, in \citet{diamond_stanic12} they argued that these outflows are most likely driven by obscured starbursts rather than by AGN. \citet{coil11} reached similar conclusions using a sample of 13 post starbursts at $0.2 < z < 0.8$. Using an unsupervised Machine Learning algorithm that searches for rare phenomena in a dataset, \citet{baron17} found a post starburst E+A galaxy with massive AGN-driven winds traced by ionized emission lines (SDSS J132401.63+454620.6). In a followup study in \citet{baron17b}, we used newly obtained ESI/Keck 1D spectroscopy to model the star formation history (SFH) and the ionized outflows in this system. In \citet{baron18} and \citet{baron20} we used the integral field units (IFUs) KCWI/Keck and MUSE/VLT to study the spatial distribution of the stars, gas, and outflows, in two post starburst galaxies hosting AGN and winds. The inferred masses and kinetic powers of the winds are 1--2 order of magnitude higher than those observed in local active galaxies. This might suggest that AGN feedback, in the form of galactic-scale outflows, is significant in the E+A galaxy phase.

Since many post starburst galaxies have undergone a recent dramatic change in their morphology and stellar population, and some of them are gas-rich (e.g., \citealt{french15, yesuf17, french18, yesuf20}), we expect their outflow properties to differ from those observed in typical active galaxies. The goal of this paper is to construct the first well-defined sample of post starburst E+A galaxies with both AGN and ionized outflows. The main difference between our sample and those presented in past studies is that the outflows in our sample are traced by strong optical emission lines, allowing us to constrain the mass and energetics of the winds (see e.g., \citealt{baron19b}). Using this sample, we aim to check whether AGN feedback, in the form of galactic outflows, can have a significant effect on their hosts evolution. In this first paper of the series, we describe the sample selection and present initial analysis of publicly-available data. In later papers we will present the analysis of a subset of this sample, observed with optical IFUs and mm observations. The paper is organized as follows. In section \ref{s:sample_selection} we describe our selection method. Section \ref{s:data_analysis} presents the data analysis, and the main results are shown and discussed in sections \ref{s:results} and \ref{s:conclusions}. We provide a description of the data availability in section \ref{s:data_avail}. Readers who are interested only in the results may skip the first two sections. Throughout this paper we assume a standard $\Lambda$CDM cosmology with $\Omega_{\mathrm{M}}=0.3$, $\Omega_{\Lambda}=0.7$, and $h=0.7$.

\section{Sample selection}\label{s:sample_selection}

Post starburst galaxies are characterized by an optical spectrum with strong Balmer absorption lines, which is dominated by A-type stars. This evolutionary stage must be very short, due to the short life-time of A-type stars, making such systems rare and comprising only $\sim$3\% of the general galaxy population (e.g., \citealt{goto03, goto07, wild09, alatalo16a}). Due to the AGN duty cycle, post starburst galaxies hosting AGN are expected to be even rarer, and post starburst galaxies hosting AGN and outflows are expected to be extremely rare. In this section we describe our method to select such sources from the \emph{Sloan Digital Sky Survey} (SDSS; \citealt{york00}) using their integrated (1D) optical spectra.

To select such rare sources, it is necessary to mine the SDSS database, which consists of more than 2 million galaxy spectra. Post starburst E+A galaxies are typically selected using either the Equivalent Width (EW) of the Balmer H$\delta$ absorption line, the D4000\AA\, index, or some combination of the two (e.g., \citealt{goto03, goto07, wild09, alatalo16a, yesuf17b, french18}). Most studies define E+A galaxies as systems that have $\mathrm{EW(H\delta) > 4-5}$\AA\footnote{In some cases additional selection criteria are used. For example, EW(H$\alpha$ emission)$< 3$\AA\, (e.g., \citealt{goto07, yesuf17b, french18}), or cuts that are based on ultraviolet and infrared colors \citep{yesuf14}.} (see however \citealt{wild09} who performed PCA decomposition of the SDSS spectra and selected sources according to both H$\delta$ and D4000\AA). Many studies use measurements provided by value added catalogs of SDSS galaxies (e.g., the MPA-JHU catalog: \citealt{b04, kauff03b, t04}; or the OSSY catalog: \citealt{oh11}). These measurements were obtained by using automatic codes. Unfortunately, these codes fail to fit the continuum emission for a non-negligible fraction of the E+A galaxies, mostly due to the stellar population templates used by the pipelines. While these templates are general enough to fit a typical SDSS galaxy well, they do not include enough young and intermediate-age stars to fit the spectrum of rare E+A galaxies. As a result, for example, some E+A galaxies in the SDSS database were classified as stars, with the code placing them at a wrong redshift of $z=0$, resulting in erroneous H$\delta$ measurement. In addition, the value added catalogs provide two estimates of the H$\delta$ absorption strength using the Lick indices \citep{worthey94}: H$\delta_{A}$ and H$\delta_{F}$. The two indices are based on different wavelength ranges, such that H$\delta_{A}$ has a wider distribution of values (from -7.5 to 10 \AA) than H$\delta_{F}$ (from -2.5 to 7 \AA). Different studies use different indices for their selection, where some select E+A galaxies as systems with H$\delta_{A} > 4-5$\AA\, (e.g., \citealt{alatalo16a, french18})\footnote{\citet{french18} include the uncertainty on the H$\delta_{A}$ index in their selection criterion, resulting in a much smaller fraction of false positives.} while others use H$\delta_{F} > 4-5$\AA\, (e.g., \citealt{yesuf14, yesuf17b}). These selection criteria result in a very different number of objects. For example, using the OSSY catalog and focusing on galaxies at redshifts 0.05--0.15, out of 453\,291 galaxies, 58\,654 have H$\delta_{A} > 5$\AA, while only 4\,760 have H$\delta_{F} > 5$\AA\, (38\,306 have H$\delta_{F} > 4$\AA). Our independent analysis and manual inspection of galaxies in the different samples suggests that selecting E+A galaxies using H$\delta_{A} > 5$\AA\, results in a non-negligible number of false positives, while selecting systems with H$\delta_{F} > 5$\AA\, misses a smaller fraction of such systems.

Our techniques to construct a sample of post starburst galaxies hosting AGN and outflows include both traditional model-fitting-based methods and Machine Learning (ML) methods. Our traditional model-fitting approach was optimized via trial and error, until the resulting sample included all the sources that were flagged as post starburst galaxies with AGN and outflows by the ML method. It is worth mentioning that none of the previous selection methods presented in the literature included \textbf{all} the objects that were flagged as post starburst galaxies by our ML algorithm. 

In our first, traditional model-fitting approach, we selected all the stellar and galaxy spectra from the SDSS. For the galaxy spectra, we focused on the redshift range $0 < z < 3$. We performed our own stellar population synthesis fit for all the objects in the initial sample, allowing the redshift to vary during the fit. For the stellar population synthesis fit, we used templates covering a wide range of ages, allowing to model different types of SFH in a non-parametric way, even that of a post starburst galaxy (see additional details in \ref{s:data_analysis:SDSS:stellar}). Using the best-fitting models, we estimated the EW of the H$\delta$ absorption line and defined E+A galaxies as systems for which $\mathrm{EW(H\delta_{F}) > 5}$\AA. After removing the stellar contribution, we analyzed the emission line spectra and performed emission line decomposition into narrow and broad kinematic components. We selected post starburst galaxies with narrow lines that are consistent with AGN photoionization (see details in \ref{s:data_analysis:SDSS:gas}). Systems with detected broad emission lines were defined as post starburst galaxies hosting AGN and ionized outflows. We filtered out type I AGN by requiring the full width half maximum (FWHM) of the broad Balmer components to be smaller than 1000 km/sec (see e.g., \citealt{shen11}). The main improvements of our method compared to other selection methods are (i) fitting the redshift and the SFH simultaneously, (ii) using more templates of young and intermediate-age stellar populations, and (iii) performing our own emission line decomposition and selecting systems with \emph{narrow lines} that are consistent with AGN photoionization. In appendix \ref{a:selection} we compare our selection method to other methods in the literature.

The second method is based on Unsupervised Machine Learning and is described in \citet{baron17}. \citet{baron17} presented a novel method to estimate distances between spectra using an Unsupervised Random Forest algorithm. This distance, which is learned from the data, traces valuable information about the properties of the population (see also \citealt{reis18}, and review by \citealt{baron19}). The distance can be used to look for objects with similar spectral properties. To select post starburst galaxies hosting AGN and ionized outflows, we applied the method to the full SDSS database and selected systems with similar spectral properties (i.e., short distances) to our patient-zero source, SDSS J132401.63+454620.6. Since this method is completely unsupervised and the distances are determined by the data, it allowed us to uncover various errors and incompletenesses in the first method, where E+A are selected by model-fitting. For reproducibility, once we have optimized our traditional model-fitting technique using the ML-selected sample, our analysis focused on the systems selected using the traditional method.

We found a total of 520 post starburst E+A galaxies with narrow emission lines that are consistent with pure AGN photoionization. Out of these, 215 show evidence for an ionized outflow in one emission line, and 144 in multiple lines. Deriving outflow properties, such as mass outflow rate and kinetic power, requires the detection of the outflow in multiple emission lines (see e.g., \citealt{baron19b}). We therefore restrict the analysis to the 144 sources for which we detected multiple broad emission lines.

\section{Data analysis}\label{s:data_analysis}

In this study we make use of publicly-available optical spectra from SDSS and far infrared photometry from the \emph{Infrared Astronomical Satellite} (IRAS; \citealt{neugebauer84}). In section \ref{s:data_analysis:IRAS} we describe our method to estimate the 60 $\mathrm{\mu m}$ flux density using IRAS, which we then use to estimate star formation rates (SFRs). In section \ref{s:data_analysis:SDSS} we describe our analysis of the optical spectra, which includes stellar population synthesis modeling (section \ref{s:data_analysis:SDSS:stellar}), emission line and absorption line fitting (section \ref{s:data_analysis:SDSS:gas}), and the estimation of AGN properties and ionized and neutral outflow energetics (sections  \ref{s:data_analysis:SDSS:AGN} and \ref{s:data_analysis:SDSS:outflow}). Finally, we describe our statistical analysis methods in section \ref{s:data_analysis:stats}.

\subsection{FIR data from IRAS}\label{s:data_analysis:IRAS}

IRAS provides a full-sky coverage in four bands centered around 12, 15, 60, and 100 $\mathrm{\mu m}$ \citep{neugebauer84}. For the 60 $\mathrm{\mu m}$ band, the positional $3 \sigma$ uncertainty for faint sources is $45'' \times 15''$. IRAS performed several scans of individual fields, and combining the information from the different scans is expected to increase the signal-to-noise of the data. To analyze the data, we used {\sc scanpi}\footnote{\url{https://irsa.ipac.caltech.edu/applications/Scanpi/}} \citep{helou88}, which is a tool for stacking the calibrated survey scans. It allows one to extract the flux of extended faint sources and to estimate upper limits for undetected sources. Our use of {\sc scanpi} to obtain FIR fluxes for nearby AGN follows the IRAS manual \citep{helou88} and the study of \citet{zakamska04}, with some modifications, and is summarized here for completeness. 

To extract the 60 $\mathrm{\mu m}$ fluxes for the galaxies in our sample, we used their SDSS coordinates. {\sc scanpi} combines all the available observations around each source and fits a point source template close to the provided coordinates. We used the default settings of {\sc scanpi}, setting the 'Source Fitting Range' to $3.2'$, the 'Local Background Fitting Range' to $30'$, and the 'Source Exclusion Range for Local Background Fitting' to $4'$. We used median stacks since the IRAS data has non-Gaussian noise. The {\sc scanpi} output includes the best-fitting flux density ($f_{\nu}$), the root mean square (RMS) deviation of the residuals after the subtraction of the best-fitting template ($\sigma$), and the offset of the peak of the best-fitting template from the specified galaxy coordinates ($\Delta$). The goodness of the fit is represented by the correlation coefficient between the best-fitting template and the data ($\rho$). For example, a point source detected with a signal-to-noise ratio of 20 has a correlation coefficient above 0.995 \citep{zakamska04}.

Since {\sc scanpi} fits a template close to the provided coordinates of each source, we expect a contamination from false matches (other sources in the $3.2' \times 3.2'$ fitting range) and from noise fluctuations. For the former we expect $\Delta$ to be large, and for the latter we expect $\rho$ to be small. Inspired by the approach presented in \citet{zakamska04}, we consider a source detected in 60 $\mathrm{\mu m}$ if the following requirements are met: $\rho > 0.8$, $\Delta < 0.4'$, and $f_{\nu}/\sigma > 3$. We selected these values using the following technique. We shuffled the right ascension (RA) values with respect to the declination (DEC) values of the sources in our sample. We then used {\sc scanpi} to stack the 60 $\mathrm{\mu m}$ observations around these shuffled coordinates. The thresholds for $\rho$, $\Delta$, and $f_{\nu}/\sigma$ were selected such that there are no detections for the shuffled coordinates, that is, no contamination from false matches or noise fluctuations. We also used a larger list of 1\,000 random coordinates and verified that the fraction of detected sources is smaller than 1\%. Using the optimized criteria, we defined the flux uncertainty of the detected sources to be $\sigma$. For the undetected sources, we defined their 60 $\mathrm{\mu m}$ flux upper limit to be $f + 3 \sigma$ and its uncertainty to be $\sigma$, where $f$ is the reported flux. We made sure that the results reported in this study do not change significantly when we change our definition of the flux upper limits (we examined a $\pm 30\%$ range around the adopted values of the flux upper limits).

Out of the 144 galaxies in our sample, 60 are detected by more than $3 \sigma$ at  60 $\mathrm{\mu m}$, 79 are upper limits, and 5 have no IRAS scans around their coordinates. We then used the measured fluxes and upper limits to estimate the star formation luminosity ($\mathrm{L_{SF}}$) and the SFR. For this we used the \citet[hereafter CE01]{chary01} templates to establish the relation between $\mathrm{\nu L_{\nu}(60\, \mu m)}$ and $\mathrm{L_{SF}}$. For the range of the observed $\mathrm{\nu L_{\nu}(60\, \mu m)}$ luminosities in our sample, the $\mathrm{L_{SF}/\nu L_{\nu}(60\, \mu m)}$ ratio of the CE01 templates is centered around 1.716, with a scatter of 0.0885 dex. We therefore defined $\mathrm{L_{SF}= 1.716 \times \nu L_{\nu}(60\, \mu m)}$ and adopted a conservative uncertainty of 0.2 dex\footnote{The uncertainty on the 60 $\mathrm{\mu m}$ flux combined with the scatter in the CE01 templates is roughly 0.15 dex.}. We estimated the SFR from $\mathrm{L_{SF}}$ using a Chabrier initial mass function (IMF; \citealt{chabrier03}) which we round off slightly to give $\mathrm{SFR = L_{SF} / \, 10^{10}\, L_{\odot}\, [M_{\odot}/yr]}$.

Many of the sources in our sample are also detected in IRAS 100 $\mathrm{\mu m}$. We ensured that the 60 $\mathrm{\mu m}$-based SFRs are consistent with those derived using 100 $\mathrm{\mu m}$. Finally, a small fraction ($< 10$) of our sources were observed by other FIR instruments, such as Herschel PACS, AKARI, or Spitzer MIPS. For these sources, we ensured that our derived SFRs are consistent with those derived by the other available FIR observations.

\subsection{Optical spectra from SDSS}\label{s:data_analysis:SDSS}

In this section we describe our analysis of the optical spectra from SDSS. In section \ref{s:data_analysis:SDSS:stellar} we describe the stellar population synthesis modeling, which was the one used to define our initial sample of galaxies. In section \ref{s:data_analysis:SDSS:gas} we describe the modeling of the emission and absorption lines, which was used to calculate the properties of the ionized and neutral gas. We then used this information to obtain the AGN (section \ref{s:data_analysis:SDSS:AGN}) and outflow properties (section \ref{s:data_analysis:SDSS:outflow}).

\subsubsection{Stellar properties}\label{s:data_analysis:SDSS:stellar}

We performed stellar population synthesis modeling using the {\sc python} implementation of \emph{Penalized Pixel-Fitting stellar kinematics extraction code} ({\sc ppxf}; \citealt{cappellari12}). This is a public code for extraction of the stellar kinematics and stellar population from absorption line spectra of galaxies \citep{cappellari04}. The code fits the input galaxy spectrum using a set of stellar libraries, finding the optimal line-of-sight velocity and velocity dispersion of the stellar population, the relative contribution of stellar populations with different ages, and the dust reddening towards the stars (assuming a \citealt{calzetti00} extinction law). For the stellar population library, we used MILES, which contains single stellar population (SSP) synthesis models and covers the full range of the optical spectrum with a resolution of full width at half maximum (FWHM) of 2.3\AA\, \citep{vazdekis10}. We used models produced with the Padova 2000 stellar isochrones assuming a Chabrier IMF. We included all the SSP models available in the library, with ages ranging from 0.03 to 14 Gyr. This large set includes many young and intermediate-age stellar populations, thus ensuring that we can properly model spectra of E+A galaxies. The output of the code includes the relative weight of stars with different ages, the line-of-sight velocity (i.e., redshift), the stellar velocity dispersion, the dust reddening towards the stars, and the best-fitting stellar model. 

To select the initial sample of the E+A galaxies presented in this study, we applied the code to all stellar and galaxy spectra (at $0 < z< 3$) from SDSS. As noted in section \ref{s:sample_selection}, some of the E+A galaxies in SDSS have a wrong redshift of $z=0$. To overcome this issue, we did not shift the observed spectra to rest-frame wavelength according to the redshifts reported by SDSS, but rather allowed the redshift to vary and used {\sc ppxf} to find the optimal redshift. For the large majority of cases, our best-fitting redshift matched the one reported by SDSS. For the few E+A galaxies that SDSS classified as stars at $z=0$, we were able to estimate the true redshift. For the 144 post starburst E+A galaxies in our final sample, we inspected the best-fitting stellar models and the resulting $\chi^2$ values, and found excellent fits in all cases. 

\subsubsection{Ionized and neutral gas properties}\label{s:data_analysis:SDSS:gas}

We obtained the emission line spectra by subtracting the best-fitting stellar models from the observed spectra. The subtracted spectra show various emission lines: [OIII]~$\lambda \lambda$ 4959,5007\AA, $\mathrm{H\alpha}$~$\lambda$ 6563\AA, [NII]~$\lambda \lambda$ 6548,6584\AA, [OII]~$\lambda \lambda$ 3725,3727\AA, $\mathrm{H\beta}$~$\lambda$ 4861\AA, [OI]~$\lambda \lambda$ 6300,6363\AA, and [SII]~$\lambda \lambda$ 6717,6731\AA\, (hereafter $\mathrm{[OIII]}$, $\mathrm{H\alpha}$, $\mathrm{[NII]}$, $\mathrm{[OII]}$, $\mathrm{H\beta}$, $\mathrm{[OI]}$, and $\mathrm{[SII]}$). We modeled each emission line as a sum of two Gaussians: one which represents the narrow component and one that represents the broader, outflowing, component\footnote{Our sample consists only of type II AGN with no broad Blamer lines from the broad line region.}. We performed a joint fit to all the emission lines under several constraints: (1) we forced the intensity ratio of the emission lines [OIII]~$\lambda \lambda$ 4959,5007\AA\, and [NII]~$\lambda \lambda$ 6548,6584\AA\, to the theoretical ratio of 1:3, (2) we tied the central wavelengths of all the narrow lines so that the gas has the same systemic velocity, and did it separately for the broad lines, and (3) we forced the widths of all the narrow lines to show the same velocity dispersion, and did the same for the broader lines. 

To select post starburst galaxies hosting AGN and outflows, we used three line-diagnostic diagrams \citep[hereafter BPT diagrams]{baldwin81, veilleux87}. The BPT diagrams are based on the line ratios [OIII]/H$\beta$, [NII]/H$\alpha$, [SII]/H$\alpha$, and [OI]/H$\alpha$, and are used to study the main source of ionizing radiation (HII regions versus AGN; see e.g., \citealt{kewley06}). To select AGN-dominated systems, we used the theoretical upper limit for SF ionized spectra by \citet{kewley01}. We defined post starburst galaxies with AGN as systems with \emph{narrow line ratios} above the theoretical upper limit in at least one of the three diagrams. For these systems, we ensured that the line ratios in the other diagrams are above or at least consistent with the theoretical threshold to within 0.2 dex. Most of the systems are classified as pure-AGN in all three diagrams, and a few systems are classified as composites in one or two diagrams. We then selected systems with broad components in H$\beta$, [OIII], [NII], and H$\alpha$, detected to at least $3 \sigma$. These form our final sample of post starburst galaxies with AGN and ionized outflows.

For our sample of 144 galaxies, we used the best-fitting parameters to estimate the flux of each emission line. We propagated the uncertainties on the best-fitting parameters to obtain the flux uncertainties. We then estimated the dust reddening towards the line-emitting gas using the measured H$\alpha$/H$\beta$ flux ratios. We assumed case-B recombination, a gas temperature of $10^4$ K, a dusty screen, and the \citet[CCM]{cardelli89} extinction law. For this extinction curve, the color excess is given by:
\begin{equation}\label{eq:1}
	{\mathrm{E}(B-V) = \mathrm{2.33 \times log\, \Bigg[ \frac{(H\alpha/H\beta)_{obs}}{2.85} \Bigg] \, \mathrm{mag} }},
\end{equation}
where $\mathrm{(H\alpha/H\beta)_{obs}}$ is the observed line ratio. We estimated the color excess for the narrow and broad components separately. We then estimated the dust-corrected line luminosities using the measured $\mathrm{E}(B-V)$. Finally, we defined the maximum blueshifted outflow velocity as $v_{\mathrm{ion \, max}} = \Delta v_{\mathrm{ion}} - 2\sigma_{\mathrm{ion}}$, where $\Delta v_{\mathrm{ion}}$ is the velocity shift of the broad lines with respect to systemic and $\sigma_{\mathrm{ion}}$ is the velocity dispersion.

To study the NaID absorption in our sources, we divided the observed spectra by the best-fitting stellar models. Since the stellar models describe the observed spectra well, particularly around the stellar MgIb absorption complex, any residual NaID absorption can be attributed to neutral gas in the interstellar medium of these galaxies (see e.g., \citealt{heckman00, rupke02}). We found evidence of NaID absorption in 69 out of 144 systems. Some of these profiles can be modeled using a single kinematic component, while others require two kinematic components to obtain a satisfactory fit. In addition, we found evidence of redshifted NaID \textbf{emission} in 21 out of 69 systems. While NaID absorption has been detected in numerous systems in the past (e.g., \citealt{rupke05a, martin06, shih10, cazzoli16, rupke17}; see a review by \citealt{veilleux20}), NaID emission has only been detected in a handful of sources (e.g., \citealt{rupke15, perna19, baron20}). 

NaID absorption is the result of absorption of continuum photons by NaI atoms along the line of sight, while resonant NaID \textbf{emission} is the result of absorption by NaI atoms outside the line of sight and isotropic reemission. For stationary NaI gas that is mixed with the stars, both absorption and emission are centered around the rest-frame wavelength of the NaID transition, resulting in net absorption. For outflowing NaI gas, the approaching part of the outflow is traced by blueshifted NaID absorption, while the receding part of the outflow emits NaID photons which are redshifted with respect to the systemic velocity. Due to their redshifted wavelength, these photons are not absorbed by NaI atoms in the stationary gas, nor by NaI atoms in the approaching side of the outflow. This results in a classical P-Cygni profile which, if observed, suggests the presence of a neutral gas outflow in the system (see additional details in \citealt{prochaska11}).

In appendix \ref{a:naid_fitting} we describe our modeling of the NaID absorption and emission profiles, and discuss possible degeneracies between the different parameters. Using the best-fitting profiles, we derived several properties of the neutral gas. We estimated the EW of the NaID absorption by integrating over the best-fitting absorption profiles. We used the best-fitting absorption optical depth $\tau_{0, K}$ to estimate the column density of the neutral Na gas, $N_{\mathrm{NaI}}$ (see equation 10 in \citealt{baron20} or \citealt{draine11}). We then estimated the hydrogen column density via (\citealt{shih10}):
\begin{equation}\label{eq:7}
	{N_{\mathrm{H}} = \frac {N_{\mathrm{NaI}}} {(1 - y) 10^{A + B+ C}}  },
\end{equation}
where $(1 - y)$ is the Sodium neutral fraction which we assume to be 0.1 (see however \citealt{baron20}), $A$ is the Sodium abundance term, $B$ is the Sodium depletion term, and $C$ is the gas metallicity term. Following \citet{shih10}, we took $A = \log [N_{\mathrm{Na}}/N_{\mathrm{H}}] = -5.69$ and $B = \log [N_{\mathrm{Na}} / N_{\mathrm{H, total}}] - \log [N_{\mathrm{Na}} / N_{\mathrm{H, gas}}] = -0.95$. For the stellar masses of the systems in our sample, the mass-metallicity relation (e.g., \citealt{t04}) suggests that the metallicity is roughly twice solar. We therefore used $C = \log [Z/ 1\, Z_{\odot}] = \log [2]$.

The best-fitting NaID absorption profiles are generally blueshifted with respect to the systemic velocity. We estimated the maximum blueshifted velocity of the NaID absorption as follows. Integrating over the total absorption profile, we defined $v_{\mathrm{NaI \, max}}$ as the velocity where the EW reaches 5\% of its total integrated value. That is, 95\% of the total EW is red-wards of $v_{\mathrm{NaI \, max}}$ and 5\% is blue-wards. We then used the best-fitting stellar velocity dispersion $\sigma_{*}$ to divide the sample into objects with stationary NaI gas versus systems with an outflow. Systems with a neutral outflow are defined as systems with $v_{\mathrm{NaI \, max}} > 2 \sigma_{*} + 70\,\mathrm{km/sec}$, where we took a safety margin of 70 km/sec which is also similar to the spectral resolution of SDSS. Out of 69 systems with NaID absorption, 58 host neutral outflows and 11 do not. We verified that all the systems showing NaID emission were classified as systems hosting an outflow according to their $v_{\mathrm{NaI \, max}}$.

\subsubsection{AGN properties}\label{s:data_analysis:SDSS:AGN}

We estimated the AGN bolometric luminosity, $\mathrm{L_{AGN}}$, using the dust-corrected narrow line luminosities of H$\beta$, [OI], and [OIII]. \citet{netzer09} presented two methods to estimate $\mathrm{L_{AGN}}$. The first method is based on the narrow H$\beta$ and [OIII] luminosities (equation 1 in \citealt{netzer09}), and the second is based on the narrow [OI] and [OIII] luminosities (equation 3). We estimated $\mathrm{L_{AGN}}$ using the two methods and found consistent results for all the objects in our sample. We adopted a conservative uncertainty of 0.3 dex for $\mathrm{L_{AGN}}$ (see \citealt{netzer09} for additional details). We then estimated the BH mass using the stellar velocity dispersion $\sigma_{*}$. The galaxies in our sample are mostly bulge-dominated. We therefore used the $M_{\mathrm{BH}}- \sigma_{*}$ relation from \citet{kormendy13} to estimate the BH mass. We assumed a nominal uncertainty of 0.3 dex since $\sigma_{*}$ is measured for the entire galaxy.

\subsubsection{Outflow energetics}\label{s:data_analysis:SDSS:outflow}

Our estimates of the ionized and neutral outflow properties are based on various earlier methods. For the ionized outflow phase, we used the prescriptions presented by \citet{baron19b}, while for the neutral outflow phase, our estimates are based on \citet{rupke05b}. Below we give a brief overview of these estimates. 

\begin{figure}
\includegraphics[width=3.5in]{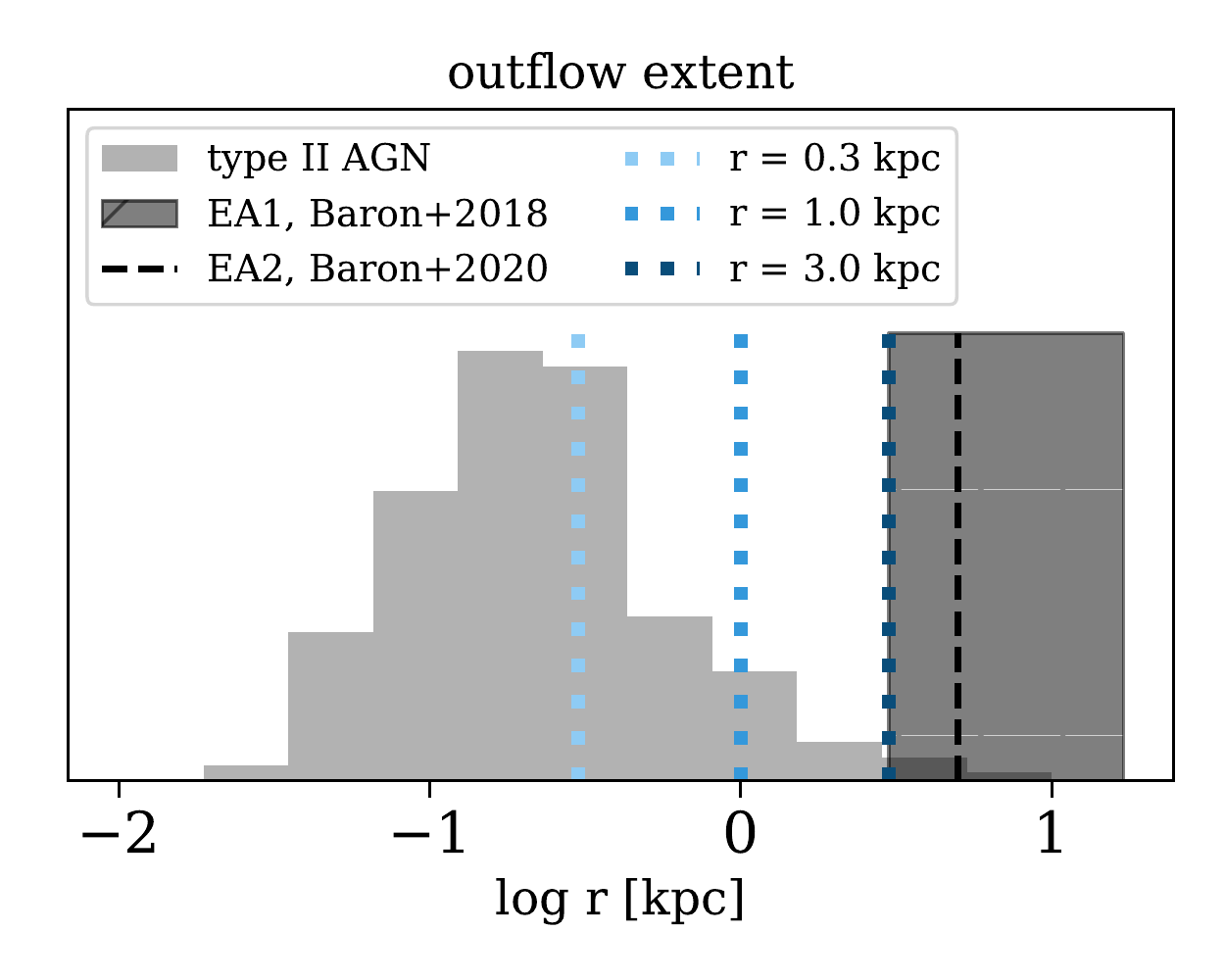}
\caption{\textbf{Comparison of our assumed ionized outflow extents to estimates from the literature.} The three assumed extents, $r=0.3$ kpc, $r=1$ kpc, and $r=3$ kpc, are marked with blue dashed vertical lines. The light grey histogram represents the distribution of ionized outflow extent derived for a large sample of type II AGN \citep{baron19a, baron19b}. The light black band represents the region where an ionized outflow is detected in the post starburst E+A galaxy SDSS J003443.68+251020.9 (hereafter EA1; \citealt{baron18}). The dashed black line represents the extent in the E+A galaxy SDSS J124754.95-033738.6 (hereafter EA2; \citealt{baron20}). Our assumed extents roughly cover the range observed in type II AGN and E+A galaxies.
}\label{f:ionized_outflow_r}
\end{figure}

To estimate the outflowing gas mass $M_{\mathrm{out}}$, mass outflow rate $\dot{M}_{\mathrm{out}}$, and kinetic power $\dot{E}_{\mathrm{out}}$ of the ionized outflow, we used equation 7 and the related text from \citet{baron19b}. These estimates require knowledge of the dust-corrected H$\alpha$ luminosity, the location of the wind, the electron density in the wind, and the effective outflow velocity. We used the dust-corrected luminosity of the broad H$\alpha$ line derived from our line fitting. The effective outflow velocity was set to $v_{\mathrm{ion \, max}}$ (see section \ref{s:data_analysis:SDSS:gas}), consistent with other estimates in the literature (e.g., \citealt{fiore17} and references therein). The spatially-integrated spectra do not allow us to estimate the outflow extent. We therefore tested three possible extents: $r=0.3$ kpc, $r=1$ kpc, and $r=3$ kpc, for all the objects in the sample. In figure \ref{f:ionized_outflow_r} we compare these assumed extents to the extent of ionized outflows in a large sample of type II AGN and in two post starburst E+A galaxies for which spatially-resolved spectroscopy is available. The three assumed extents roughly cover the range of extents observed in type II AGN and E+A galaxies. 

We used the ionization parameter method presented in \citet{baron19b} to estimate the electron density in the outflow (equation 4 there). This method is based on the relation between the AGN luminosity, the gas location, and its ionization state. It requires knowledge of $\mathrm{L_{AGN}}$, the outflow location, and the ionization parameter. We used the broad emission line ratios [NII]/H$\alpha$ and [OIII]/H$\beta$ to estimate the ionization parameter in the outflowing gas (equation 2 in \citealt{baron19b}). This resulted in three estimates of the electron density in the outflowing gas, corresponding to the three assumed outflow extents. The advantage of this method is that the derived electron density and the assumed outflow extent are consistent with the observed line ratios assuming AGN-photoionized gas. This is different from many cases in the literature, where the combination of assumed/derived electron densities and derived outflow extents is inconsistent with the observed line ratios (see discussion in \citealt{baron19b, davies20, santoro20}).

The derived $M_{\mathrm{out}}$, $\dot{M}_{\mathrm{out}}$, and $\dot{E}_{\mathrm{out}}$ are highly uncertain and should not be studied individually. However, the true outflow properties are  expected to be within the range spanned by the three sets of ($M_{\mathrm{out}}$, $\dot{M}_{\mathrm{out}}$, $\dot{E}_{\mathrm{out}}$)$_{r=0.3, 1, 3\,\mathrm{kpc}}$. In section \ref{s:results:outflow_properties} we use the three sets to perform a qualitative comparison of the outflow properties in post starburst E+A galaxies with those observed in other systems (type II AGN and ULIRGs). Finally, these properties were derived using estimates of the electron density, which in turn depends on $\mathrm{L_{AGN}}$ through the ionization parameter method. Therefore, we expect these properties to be correlated with $\mathrm{L_{AGN}}$. In section \ref{s:results:outflow_properties:driver} we examine the main driver of the observed outflows, by looking for correlations between the outflow properties and $\mathrm{L_{AGN}}$ and $\mathrm{L_{SF}}$. To avoid this dependency, we estimated ($M_{\mathrm{out}}$, $\dot{M}_{\mathrm{out}}$, $\dot{E}_{\mathrm{out}}$) under the assumption of $n_e = 10^3\,\mathrm{cm^{-3}}$ and used these in section \ref{s:results:outflow_properties:driver}. Assuming $r = 1$ kpc, this combination of electron density and outflow extent is roughly consistent with the observed line ratios in our sources.

To estimate $M_{\mathrm{out}}$, $\dot{M}_{\mathrm{out}}$, and $\dot{E}_{\mathrm{out}}$ in the neutral wind, we followed \citet{rupke05b} and assumed the time-averaged thin-shell model. We used the following expressions:
\begin{equation}\label{eq:8}
	{M_{\mathrm{out}} = 5.6 \times 10^7 \, M_{\odot}\, \Big( \frac{C_{\Omega}}{0.4} C_f\Big) \Big( \frac{N_{\mathrm{H}}}{10^{21} \, \mathrm{cm^{-2}}}\Big) \Big( \frac{r}{1\,\mathrm{kpc}}\Big)^2},
\end{equation}
\begin{equation}\label{eq:9}
	\begin{split}
	& \dot{M}_{\mathrm{out}} = 11.45 \, M_{\odot}/\mathrm{yr}\, \Big( \frac{C_{\Omega}}{0.4} C_f\Big) \Big( \frac{N_{\mathrm{H}}}{10^{21} \, \mathrm{cm^{-2}}}\Big) \Big( \frac{r}{1\,\mathrm{kpc}}\Big) \times \\
	& \,\,\,\,\,\,\,\,\,\,\,\,\,\,\,\,\,\,\,\, \Big( \frac{v}{200\,\mathrm{km/sec}}\Big),
	\end{split}
\end{equation}
\begin{equation}\label{eq:10}
	{\dot{E}_{\mathrm{out}} = 3.2 \times 10^{35} \, \mathrm{erg/sec}\, \Big( \frac{\dot{M}_{\mathrm{out}}}{M_{\odot}/\mathrm{yr}} \Big) \Big( \frac{v}{\mathrm{km/sec}}\Big)^2},
\end{equation}
where $C_{\Omega}$ is the the large-scale covering factor related to the opening angle of the wind, and $C_f$ is the local covering fraction related to its clumpiness. $N_{\mathrm{H}}$ is the hydrogen column density, $r$ is the wind location, and $v$ is the outflow velocity. These expressions are equivalent to equations 13, 14, and 18 in \citet{rupke05b}, except for a factor 10 introduced to correct a typo in their equations 13--18.

We estimated $M_{\mathrm{out}}$, $\dot{M}_{\mathrm{out}}$, and $\dot{E}_{\mathrm{out}}$ in the 58 systems in which we detected neutral outflows using the best-fitting $C_f$ and the derived $N_{\mathrm{H}}$ values (equation \ref{eq:7}). We defined the wind velocity to be the maximum blueshifted NaID velocity, $v_{\mathrm{NaI \, max}}$. To estimate the large-scale covering factor $C_{\Omega}$, we assumed that the detection rate of neutral outflows in post starburst galaxies with AGN is related to the wind opening angle. Out of all the post starburst galaxies with AGN, roughly 20\% show signatures of a neutral outflow. We therefore took $C_{\Omega} = 0.2$. Similar to the ionized outflow case, we assumed three outflow extents: $r=0.3$ kpc, $r=1$ kpc, and $r=3$ kpc. 

The estimated $M_{\mathrm{out}}$, $\dot{M}_{\mathrm{out}}$, and $\dot{E}_{\mathrm{out}}$ of the neutral gas are also highly uncertain. We use the three sets of neutral ($M_{\mathrm{out}}$, $\dot{M}_{\mathrm{out}}$, $\dot{E}_{\mathrm{out}}$)$_{r=0.3, 1, 3\,\mathrm{kpc}}$ to perform a qualitative comparison with neutral outflow properties presented in the literature (section \ref{s:results:outflow_properties}). The mass and energetics of the ionized and neutral outflows depend on the assumed location $r$ in a similar manner. Assuming that the neutral and ionized outflows have roughly similar extents, the neutral-to-ionized gas mass, mass outflow rate, and kinetic power do not depend on the assumed location. In section \ref{s:results:outflow_properties:multiphased} we use this assumption to study the multi-phased nature of the wind, and compare the neutral-to-ionized ($M_{\mathrm{out}}$, $\dot{M}_{\mathrm{out}}$, $\dot{E}_{\mathrm{out}}$) ratios to those presented in the literature. 

Our adopted uncertainties on $M_{\mathrm{out}}$, $\dot{M}_{\mathrm{out}}$, and $\dot{E}_{\mathrm{out}}$ of the neutral and ionized outflows include only those associated with the line-fitting process, propagated through the different equations described in this section. They do not include the uncertain wind geometry, or other uncertain gas properties such as the metallicity or the Sodium neutral fraction (see \citealt{harrison18} and \citealt{veilleux20} for more details). However, with find the uncertainty associated with the unknown outflow location to be the dominant factor, and we account for it by presenting ranges of $M_{\mathrm{out}}$, $\dot{M}_{\mathrm{out}}$, and $\dot{E}_{\mathrm{out}}$ for the three assumed extents (see figures \ref{f:ionized_outflow_M_and_E_comparison} and \ref{f:neutral_outflow_M_and_E_comparison_ver2} below).

\subsection{Statistical analysis}\label{s:data_analysis:stats}

Our sample includes a significant number of upper limits on $\mathrm{L_{SF}}$. We therefore used \emph{Survival Analysis} techniques to study distributions and correlations between different properties (see \citealt{feigelson85, isobe86} and references within). In particular, we used the Kaplan-Meier Estimator (KME) to derive cumulative distribution functions (CDFs) for censored datasets (i.e., datasets with upper limits) and the logrank test to test whether two censored samples are drawn from the same parent distribution (see \citealt{isobe86}). For these two, we restricted the analysis to datasets with detection fractions larger than 10\%. We used the generalized Kendall's rank correlation to estimate the correlation between properties with upper limits. For the KME and logrank test, we used the Python library {\sc lifelines} (version 0.25.4; \citealt{davidson_pilon2020}). For the generalized Kendall's rank correlation, we used {\sc pymccorrelation}\footnote{\url{https://github.com/privong/pymccorrelation}}, first presented in \citet{privon20}. We used the Python implementation\footnote{\url{https://github.com/jmeyers314/linmix}} of {\sc linmix} to perform linear regression. {\sc linmix} is a Bayesian method that accounts for both measurement uncertainties and upper limits \citep{kelly07}. 

\section{Results and Discussion}\label{s:results} 

\begin{table*}

 \caption{\textbf{Far infrared properties of the galaxies in the different samples.} The first four rows represent samples of post starburst E+A galaxies with different emission line properties. These samples are ordered according to the mean luminosity of the narrow emission lines in a descending order. The last row represents typical AGN host galaxies from SDSS. The table lists the fractions of FIR-detected sources and upper limits. It also lists the fraction of galaxies we consider to be above the star forming main sequence ($> 0.3$ dex) and the fraction of starbursts ($> 0.6$ dex), with their associated confidence intervals for confidence level of 95\%. The table shows that post starburst galaxies with stronger emission lines tend to be more luminous in FIR, with a larger fraction of galaxies above the main sequence. Among AGN host galaxies, post starburst E+A galaxies are more likely to be above the main sequence compared to typical AGN host galaxies. }\label{tab:detection_frac}
 	\centering
\begin{tabular}{|m{2cm} | m{1.5cm} m{3.3cm} c| c c | m{2cm} | m{2cm}| } 
\hline
Sample & \multicolumn{2}{l}{Selection criteria}  & $\mathrm{N_{obj}}$ & \multicolumn{2}{c|}{IRAS 60 $\mathrm{\mu m}$ properties$^{\dagger}$} & \% above the MS ($>0.3$ dex) & \% of starbursts ($>0.6$ dex)  \\
       & Galaxy type & Emission line properties   &                    & \# detections & \# upper limits &  &  \\
\hline
\hline
This work & E+A & AGN photoionization \& outflows: including LINERs, Seyferts, and composites. & 144 & 60 (41\%) & 79 (54\%) & 45\% (35--56\%) & 32\% (24--41\%)  \\
  \hline
\citet{yesuf17b} & E+A & AGN photoionization: only Seyferts, $\sim$30\% of the objects show outflows. & 524 & 135 (25\%) & 366 (70\%) & 30\% (25--35\%) & 18\% (15--22\%) \\ 
  \hline
\citet{alatalo16a} & E+A & AGN/SF photoionization: emission line ratios consistent with shocks. & 650 & 84 (13\%) & 538 (83\%) & 13\% (10--17\%) & 7\% (5--9\%) \\
 \hline
\citet{french18} & E+A & weak or no emission lines: EW(H$\alpha$) $< 3$\AA. & 532 & 24 (4.5\%) & 488 (92\%) & - & -  \\
  \hline
  \hline
Type II AGN & $z \sim 0.1$ galaxy & AGN photoionization: only Seyferts. & 6721 & 1892 (28\%) & 4829 (72\%) & 16\% (15--17\%) & 6\% (5.3--6.5\%) \\
\hline

\multicolumn{8}{l}{$^{\dagger}$ The fraction of detected sources and upper limits do not add up to 100\% since there are sources for which no IRAS scans are available.} \\

\end{tabular}
\end{table*}

We constructed a sample of post starburst galaxy candidates with AGN and ionized outflows. We found that $\sim$40\% of the post starburst galaxies with AGN show evidence of ionized outflows. For comparison, $\sim$50\% of type II AGN at $0.05 < z < 0.15$ show evidence of ionized outflows (see e.g., \citealt{mullaney13, baron19a}). We concentrated on a subset of the sample in which we detected the outflow in multiple emission lines. We used FIR observations by IRAS to estimate their SF luminosities $\mathrm{L_{SF}}$ or their upper limits. We performed stellar population synthesis modeling of their optical spectra, and obtained their best-fitting SFH, stellar velocity dispersion, and dust reddening towards the stars. We decomposed the observed emission lines into narrow and broad kinematic components, which correspond to stationary NLR and outflowing gas respectively. Using these properties, we estimated the AGN bolometric luminosity, $\mathrm{L_{AGN}}$, and the BH mass, $M_{\mathrm{BH}}$. We detected NaID absorption in 69 systems, 21 of which show evidence of NaID \textbf{emission}. The NaID emission and absorption trace neutral gas in the ISM of these galaxies. We modeled the NaID line profiles and found that 58 out of the 69 systems host neutral gas outflows. Using the derived ionized and neutral gas properties, we presented rough estimates of the mass, mass outflow rate, and kinetic power in the ionized and neutral winds. 

In this section we present and discuss our main results. In section \ref{s:results:E_A_are_star_forming} we use the FIR-based SFRs to show that many post starburst galaxy candidates, in our sample and in other samples from the literature, actually host starbursts that are highly obscured in optical wavelengths. We also rule out other scenarios as major contributors to the FIR. We then study the relation between $\mathrm{L_{SF}}$ and various system properties in section \ref{s:results:SF_correlations}. In section \ref{s:results:outflow_properties} we study the ionized and neutral outflow properties in our sources.

\subsection{Many "post starburst" galaxies host obscured starbursts}\label{s:results:E_A_are_star_forming} 

Post starburst galaxy candidates have been typically selected through their optical spectra. As a result, their SF properties were derived from optical observations. Traditionally, quenched post starburst E+A galaxies have been selected to have strong H$\delta$ absorption and weak emission lines (see e.g., \citealt{yesuf14, alatalo16a, french18} and references therein). However, selection of systems with weak emission lines misses transitioning post starburst galaxies in which residual SF ionizes the gas, or systems with other ionization sources, such as AGN or shocks (\citealt{yesuf14, alatalo16a}). To include transitioning post starbursts in their selection, \citet{yesuf14} used a combination of different observables, and identified an evolutionary sequence from starbursts to fully quenched post-starbursts. While their scheme provides a more complete picture of post starburst galaxies, it does not properly account for two important sub-classes: AGN-dominated systems and heavily-obscured starbursts. In particular, they define the location of a galaxy within the sequence according to the EW of the H$\alpha$ emission. Thus, systems that host obscured starbursts can be misclassified within their sequence. For example, \citet{poggianti00} found that $\sim$50\% of their very luminous infrared galaxies show E+A-like signatures. These obscured starbursts would have been classified as quenched post starburst galaxies by both the traditional and the \citet{yesuf14} schemes (for additional examples see \citealt{tremonti07, diamond_stanic12}).  

In this section we use the FIR observations by IRAS to study the SF properties of post starburst galaxy candidates. We show that some of these are in fact obscured starbursts, similar to the objects presented by \citet{poggianti00}. For these sources, optical-based derived properties, such as the H$\alpha$ emission EW, the Dn4000\AA\, index, or the age of the stellar population (as inferred from the optical spectrum; e.g., \citealt{french18}), do not fully represent their nature. These objects would have been misclassified by the scheme of \citet{yesuf14}. 

To study these issues, we used several galaxy samples (see details in appendix \ref{a:other_samples}). We used four different samples of post starburst galaxy candidates with different emission line properties. Some were selected to host AGN and/or outflows, while others were selected to have weak or no emission lines (i.e., traditional E+A galaxies). For comparison, we also used a sample of AGN from SDSS. In table \ref{tab:detection_frac} we summarize the properties of the different samples, including their selection criteria, their FIR detection fractions, and the fraction of galaxies above the SFMS (see below). The galaxies in the different samples have somewhat different distributions of stellar mass and redshift. Therefore, whenever we compare between the SF properties of two samples, we match their stellar masses and redshifts and ensure that the conclusions do not change. 

\begin{figure}
\includegraphics[width=3.5in]{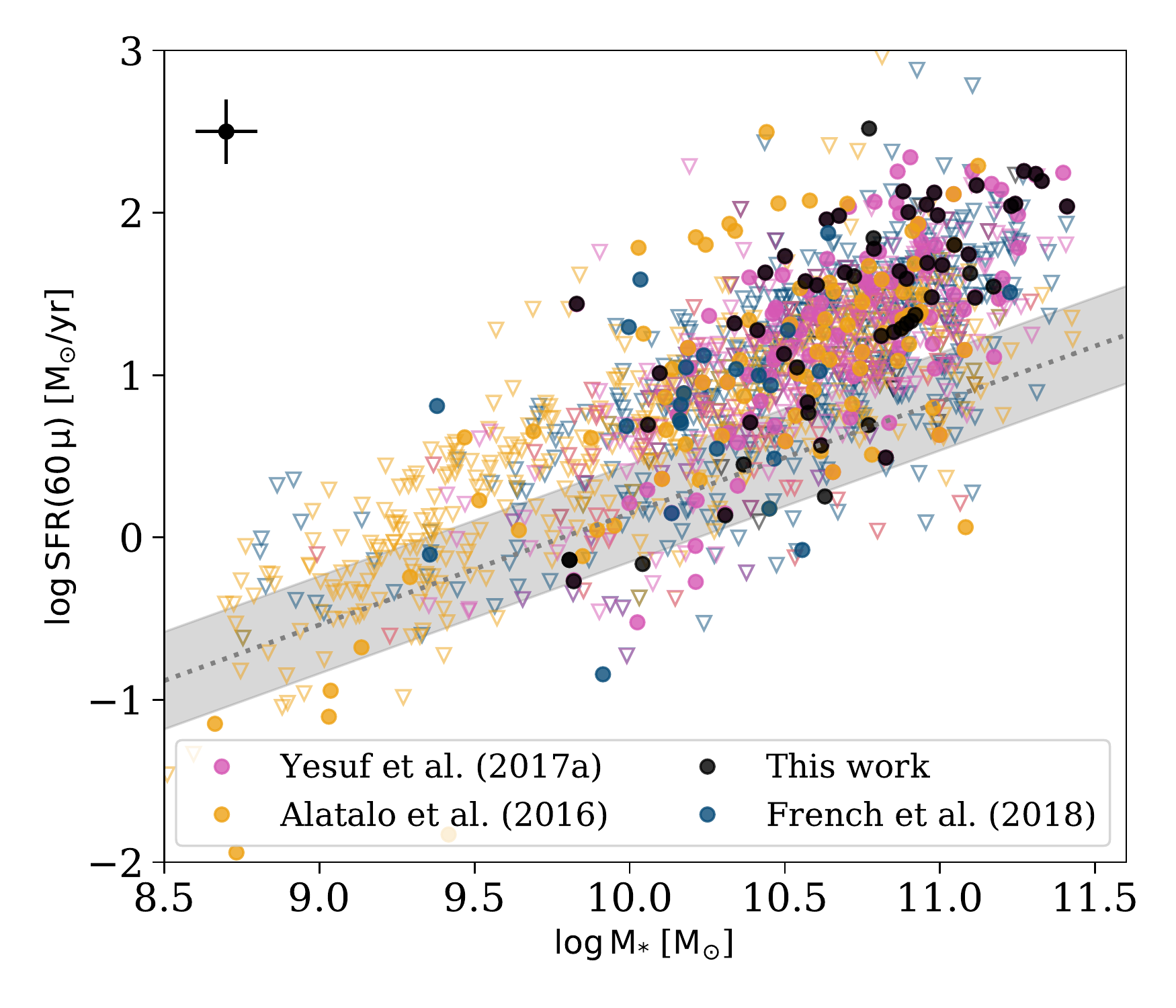}
\caption{\textbf{SFR versus stellar mass for post starburst galaxy candidates from different studies.} 
The circles represent the 60 $\mathrm{\mu m}$-based SFRs of the post starburst galaxies in the different samples, and the empty triangles represent the upper limits. The marker colors correspond to different samples, where our sample is marked with black, \citet{french18} with dark-blue, \citet{alatalo16a} with orange, and \citet{yesuf17b} with pink. The typical uncertainty of the measurements is marked at the top left part of the diagram. The grey dotted line represents the SFMS at $z = 0.1$ \citep{whitaker12}, and the light-grey band a $\pm 0.3$ dex interval around it. Most of the FIR-detected galaxies are located on or above the SFMS.
}\label{f:SFR_vs_M_all_samples}
\end{figure}

Table \ref{tab:detection_frac} lists the fraction of sources detected in 60 $\mathrm{\mu m}$ in the different samples. Given the flux limit of IRAS ($\sim$ 0.2 Jy at 60 $\mathrm{\mu m}$) and the redshifts of the galaxies ($z \sim 0.1$), the large detection fractions suggest that many of these systems are in fact forming stars at this stage. In figure \ref{f:SFR_vs_M_all_samples} we show the location of the galaxies on the SFR-$M_{*}$ diagram using the 60 $\mathrm{\mu m}$-based SFRs, and the stellar masses provided by the MPA-JHU catalog \citep{b04, kauff03b, t04}. We compare their location to the SFMS at $z = 0.1$ as presented in \citet{whitaker12}. Most of the FIR-detected sources lie on or above the SFMS.

\begin{figure}
	\centering
\includegraphics[width=3.5in]{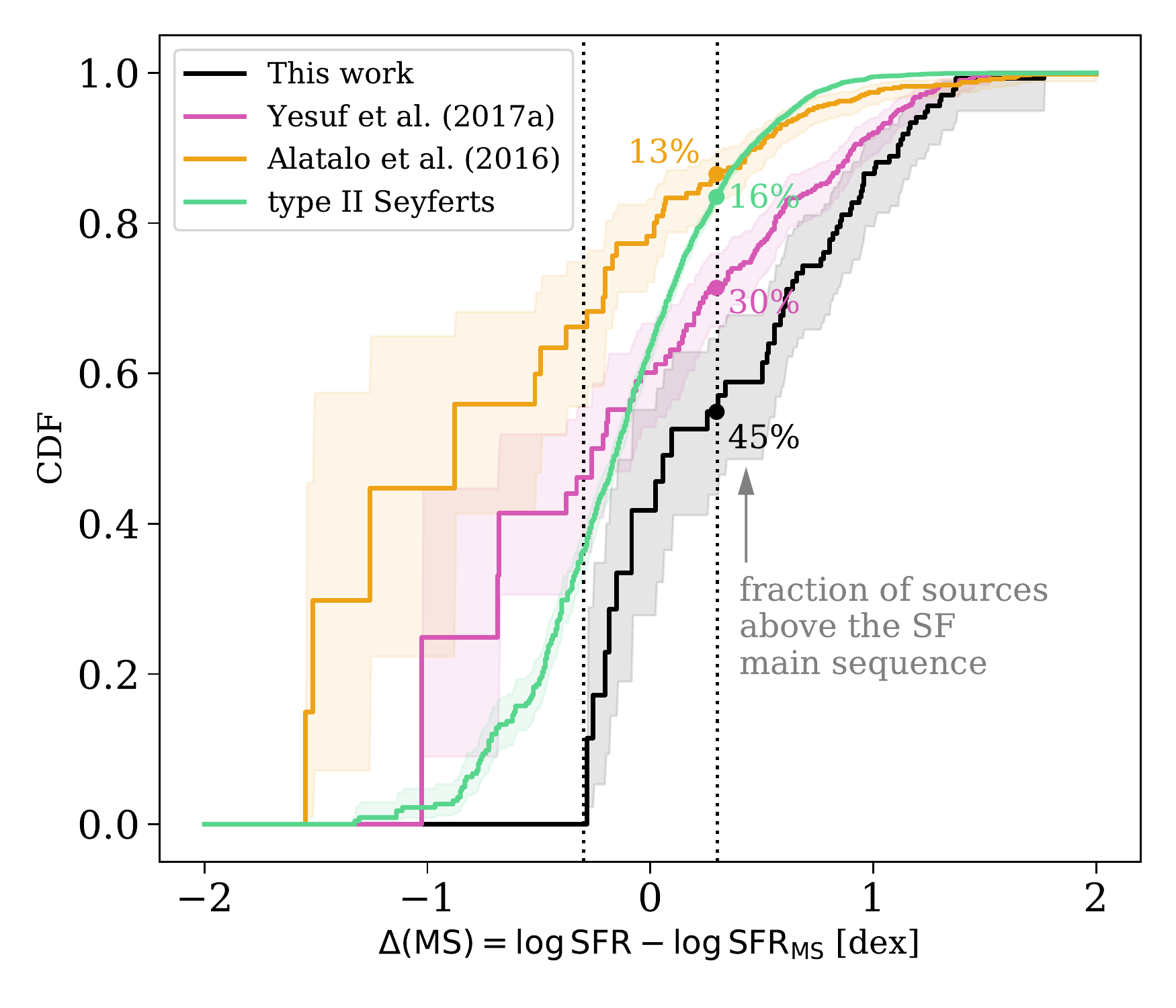}
\caption{\textbf{Cumulative distribution functions (CDFs) of $\Delta(\mathrm{MS}) =  \log \mathrm{SFR} - \log \mathrm{SFR_{MS}}$ of the galaxies in the different samples.} 
Each solid line represents a CDF, estimated by using the Kaplan-Meier Estimator. The transparent bands mark the confidence intervals of each CDF (confidence level of 95\%). 
The colors represent different samples as indicated in the diagram. The vertical dotted lines represent $\Delta(\mathrm{MS}) = \pm 0.3$ dex. For each sample, we list the inferred fraction of sources above the SFMS.
 }\label{f:Delta_MS_all_samples}
\end{figure}

To include the upper limits in the analysis, we define the distance from the main sequence $\Delta(\mathrm{MS}) = \log \mathrm{SFR} - \log \mathrm{SFR(M_{*}, z)_{MS}}$ for all the sources. The second term represents the expected SFR of a main sequence galaxy with stellar mass $M_{*}$ at redshift $z$ (hereafter $\log \mathrm{SFR_{MS}}$). We define a galaxy to be above the main sequence if $\Delta(\mathrm{MS}) > 0.3$ dex. In figure \ref{f:Delta_MS_all_samples} we show the CDFs of $\Delta(\mathrm{MS})$ for the different samples. These were calculated using the KME, which accounts for upper limits in the data. According to the estimated CDFs, 45\% of the galaxies in our sample are above the main sequence, with confidence intervals 35--56\% (confidence level of 95\%). For the two other post starburst candidate samples with emission lines, the fractions of sources above the main sequence are 30\% and 13\% (see table \ref{tab:detection_frac})\footnote{We used the linear star forming main sequence from \citet{whitaker12}. \citet{saintonge16} suggested a non-linear star forming main sequence which flattens at higher stellar masses. Adopting the \citet{saintonge16} main sequence results in a larger fraction of sources above the main sequence which strengthens our conclusions.}. The sample by \citet{french18}, which was selected to consist of quenched post starburst galaxies, has the smallest FIR detection fraction of 4.5\%, and was therefore excluded from this analysis. Nevertheless, figure \ref{f:SFR_vs_M_all_samples} shows that most of the FIR-detected objects in this sample lie above the SFMS as well. 

We define a galaxy as a starburst if it is located 0.6 dex above the SFMS (see e.g., \citealt{rodighiero11}). According to the estimated CDFs, 32\% of the galaxies in our sample are starbursts. For the two other post starburst candidate samples with emission lines, the fractions of starbursts are 18\% and 7\%. These fractions suggest that many systems that were classified as transitioning post starburst galaxies are in fact starbursts. These conclusions are in line with the findings of \citet{poggianti00}.

\begin{figure}
	\centering
\includegraphics[width=3.5in]{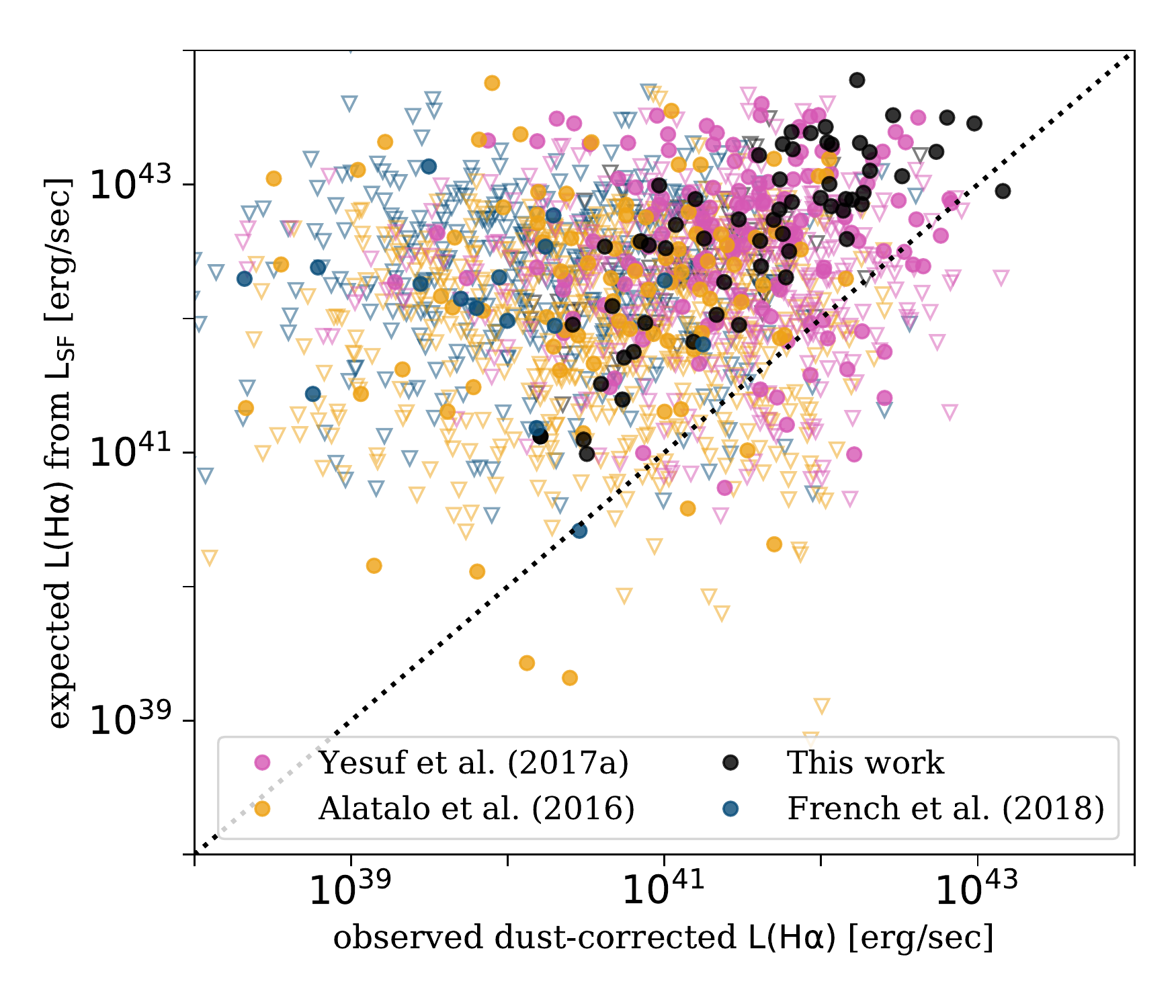}
\caption{\textbf{Comparison of SFR indicators in FIR and optical.}
The expected H$\alpha$ luminosity derived from the FIR-based SFR versus the observed dust corrected H$\alpha$ luminosity. The marker colors and shapes are similar to those of figure \ref{f:SFR_vs_M_all_samples}. The black dashed line represents a 1:1 line. The objects in our sample and in the Yesuf et al. sample are AGN-dominated (Seyferts and LINERs). Therefore, the observed H$\alpha$ emission in these samples is dominated by the AGN and not by the SF. The large majority of the FIR-detected sources lie above the 1:1 line, suggesting that the star formation is obscured.
 }\label{f:Halpha_expected_versus_observed}
\end{figure}

Next, we examine whether these starbursts are obscured. We use the SF luminosity to estimate the H$\alpha$ luminosity that would have been observed if the SF was not obscured. Assuming a Charier IMF, $\mathrm{L(H\alpha) \approx L_{SF} / 200}$ (see e.g., \citealt{calzetti13}). In figure \ref{f:Halpha_expected_versus_observed} we show the expected H$\alpha$ luminosity versus the observed dust-corrected H$\alpha$ luminosity. The large majority of the FIR-detected sources lie above the 1:1 line, suggesting that the observed H$\alpha$ underestimates the true SF in these systems. Considering only the FIR-detected sources, the median L(H$\alpha$ expected)/L(H$\alpha$ observed) is 9, 15, 28, and 230, for our sample, Yesuf et al., Alatalo et al., and French et al. samples, respectively. The objects in our sample and in the Yesuf et al. sample are AGN-dominated (Seyfert and LINERs), thus their H$\alpha$ emission is dominated by the AGN and not by SF. According to \citet{wild10}, for AGN-dominated systems, the median contribution of the SF to the H$\alpha$ emission is 20\%. This suggests that the median L(H$\alpha$ expected)/L(H$\alpha$ observed) in these two samples is at least 45 and 75. To summarize, the optical-based SFR is 1--2 magnitudes smaller than the FIR-based SFR, suggesting that these starbursts are highly obscured\footnote{The H$\alpha$ luminosity is the integrated H$\alpha$ emission within the SDSS 3'' fiber, while the FIR-based SFR is a global measure for the entire galaxy. However, for the typical redshift in our sample, $z \sim 0.1$, the SDSS fiber covers roughly 6 kpc of the galaxy. Therefore, we find it unlikely that the difference between the FIR-based SFR and optical-based SFR is due to fiber effects.}. This also suggests that other optically-derived properties, such as the Dn4000\AA\, index and the age of the stellar population, do not represent the true SF nature in these systems. These conclusions are also in line with the findings of \citet{poggianti00}.

Table \ref{tab:detection_frac} suggests that post starburst candidates with more luminous emission lines tend to have a larger fraction of starbursts. For each sample, we estimated the dust-corrected H$\alpha$ luminosity, and ordered the samples in the table according to the median luminosity in a descending order. The table also shows that both the FIR detection fraction and the fraction of galaxies above the main sequence decrease from one row to the next. In section \ref{s:results:SF_correlations} below, we also find a significant correlation between $\mathrm{L_{SF}}$ and the narrow and broad H$\alpha$ luminosities in our sample.

Finally, we show that among AGN hosts, post starburst candidates are more likely to be found above the SFMS. For that, we used the Yesuf et al. and type II AGN samples. The two samples have similar emission line properties (Seyfert-like emission line ratios), but they differ in their optically-based stellar selection criteria. In particular, the objects in the Yesuf et al. sample are post starburst candidates with EW(H$\delta$) $> 4$ \AA. We matched the samples to have similar distributions in redshift, stellar mass, and Dn4000\AA\, index. We then used survival analysis to calculate the CDFs of $\Delta(\mathrm{MS})$ for the matched samples. We find that 34\% (28\% -- 41\%) of the post starburst Seyferts are above the SFMS, compared to only 16\% (12\% -- 20\%) in the general Seyfert population. The fractions of starbursts are 21\% (17\% -- 26\%) and 6\% (4\% -- 9\%) respectively. This suggests that systems selected to have post starburst signatures in the optical (EW(H$\delta$) $> 4$ \AA) are more likely to be starbursts.

There are several caveats to using the FIR luminosity to estimate the SFR of a post starburst system (e.g., \citealt{kennicutt98}). First, the FIR emission traces the SFR averaged over the past $\sim$100 Myrs. Therefore, if the SFR varied significantly during the past 100 Myrs, the FIR-based SFR may not represent the instantaneous SFR of the system. In addition, older stars can heat the dust as well. While there are no O or B-type stars in post starburst galaxies, there may be enough A-type stars to heat the dust and produce a significant FIR emission. \citet{hayward14} used hydrodynamical simulations of galaxy mergers to show that the FIR luminosity can significantly overestimate the instantaneous SFR in post starburst galaxies. This, however, applies for systems with SFR(FIR) $\lesssim1 \, \mathrm{M_{\odot}/yr}$, whereas our FIR-detected sources have SFR(FIR) $\gtrsim 10 \, \mathrm{M_{\odot}/yr}$. In this regime of SFR(FIR) $\gtrsim 10 \, \mathrm{M_{\odot}/yr}$, the simulations by \citet{hayward14} show a rough agreement between SFR(FIR) and the instantaneous SFR. 

To further test the possibility that the FIR emission is driven by A-type stars, we used {\sc starburst99} \citep{leitherer99} to simulate the SED of a stellar population following an instantaneous starburst that produced $M = 10^{7}\,\mathrm{M_{\odot}}$ (all the other parameters were set to their default values). We executed {\sc starburst99} to $t=10^{9}$ (yr)\footnote{After 1 Gyr, the bolometric luminosity of the stellar population is negligible compared to the luminosity of stars produced in more recent episodes.}. We scaled the {\sc starburst99} SEDs according to the derived SFH and calculated the expected SED of each post starburst galaxy. We estimated the total bolometric luminosity of the stellar population $L_{\mathrm{SF}}^{\mathrm{bol}}$ by integrating over the SED. In the most extreme case, the dust covers the stellar population entirely and absorbs all the radiation originating from it. In this case, the dust FIR emission $L_{\mathrm{dust}}^{\mathrm{FIR}}$ equals to $L_{\mathrm{SF}}^{\mathrm{bol}}$. In the more realistic case of partial obscuration, $L_{\mathrm{dust}}^{\mathrm{FIR}} < L_{\mathrm{SF}}^{\mathrm{bol}}$. For most of the FIR-detected sources, the derived $L_{\mathrm{SF}}^{\mathrm{bol}}$ is $\sim 10$ smaller than the observed FIR emission. In addition, we used the best-fitting stellar SEDs and the derived dust reddening values from section \ref{s:data_analysis:SDSS:stellar} to calculate the unobscured bolometric luminosity of the stellar population. These luminosities are 10--100 times smaller than the dust FIR luminosities. All these suggest that the transitioning post starburst observed in the optical spectra did not produce enough A-type stars to power the observed FIR emission. It strengthens the conclusion that the FIR emission is due to an ongoing, heavily obscured, SF.

\subsubsection{Alternative Scenarios}\label{s:results:E_A_are_star_forming:alternative}

In the above discussion, we assumed that the FIR emission originates from the interstellar medium and is powered by SF. In this section we discuss alternative scenarios and demonstrate that they are unlikely.

The observed FIR emission can be emission from dust that is mixed with the ionized or neutral outflow (see e.g., \citealt{baron19a}). In this scenario, the dust that is mixed with the outflow is heated by the AGN, and then emits in IR wavelengths. To test this scenario, we estimated the dust mass from the observed 60 $\mathrm{\mu m}$ luminosity using the modified black body method summarized in \citet{berta16}, and using the dust opacity by \citet{li01}. We assumed a dust temperature of 25--30 (K), which is consistent with the temperatures expected for galaxies with similar SF luminosities (see \citealt{magnelli14}). Assuming gas to dust ratio of $\sim$100, the resulting gas mass is at least one order of magnitude larger than the estimated outflowing gas mass in the ionized and neutral phases (see section \ref{s:results:outflow_properties} for additional details). It therefore excludes the possibility that the FIR emission originates from dust that is mixed with the wind.

Finally, we demonstrate that the $60\,\mathrm{\mu m}$ emission can not be powered by the AGN. The intrinsic AGN SED at infrared wavelengths has been the subject of extensive research (see \citealt{lani17} and references therein). Both observations and clumpy torus models show that the intrinsic AGN SED drops significantly above $20\,\mathrm{\mu m}$, with the luminosity ratio $\nu L_{\nu}(60\,\mathrm{\mu m}) / \nu L_{\nu}(22\,\mathrm{\mu m})$ ranging from 0.17 to 0.3 (e.g., \citealt{mor09, stalevski12, lani17}). In contrast, the FIR-detected post starbursts in the different samples show $\nu L_{\nu}(60\,\mathrm{\mu m}) / \nu L_{\nu}(22\,\mathrm{\mu m}) \sim 3$, consistent with the ratio observed in star forming galaxies \citep{chary01}. The objects which are not AGN-dominated (weak emission lines or classified as star forming) show mid--far infrared SEDs that are well-fitted with the \citet{chary01} templates of star forming galaxies. The AGN-dominated systems (Seyferts or LINERs) are well-fitted with a combination of a torus and a SF template, where the former dominates at mid infrared while the later dominates at far infrared wavelengths (see \citealt{baron19a} for details about the SED fitting). Even in the most extreme cases where the AGN dominates the $22\,\mathrm{\mu m}$ completely, its contribution to the $60\,\mathrm{\mu m}$ flux is negligible. 

\subsection{Relations between SF and other system properties}\label{s:results:SF_correlations}

In this section we present the relations between $\mathrm{L_{SF}}$ and various other system properties. In particular, we examine the relation with the AGN luminosity, with optical-based stellar properties, and with the gas properties. 

\begin{figure}
	\centering
\includegraphics[width=3.5in]{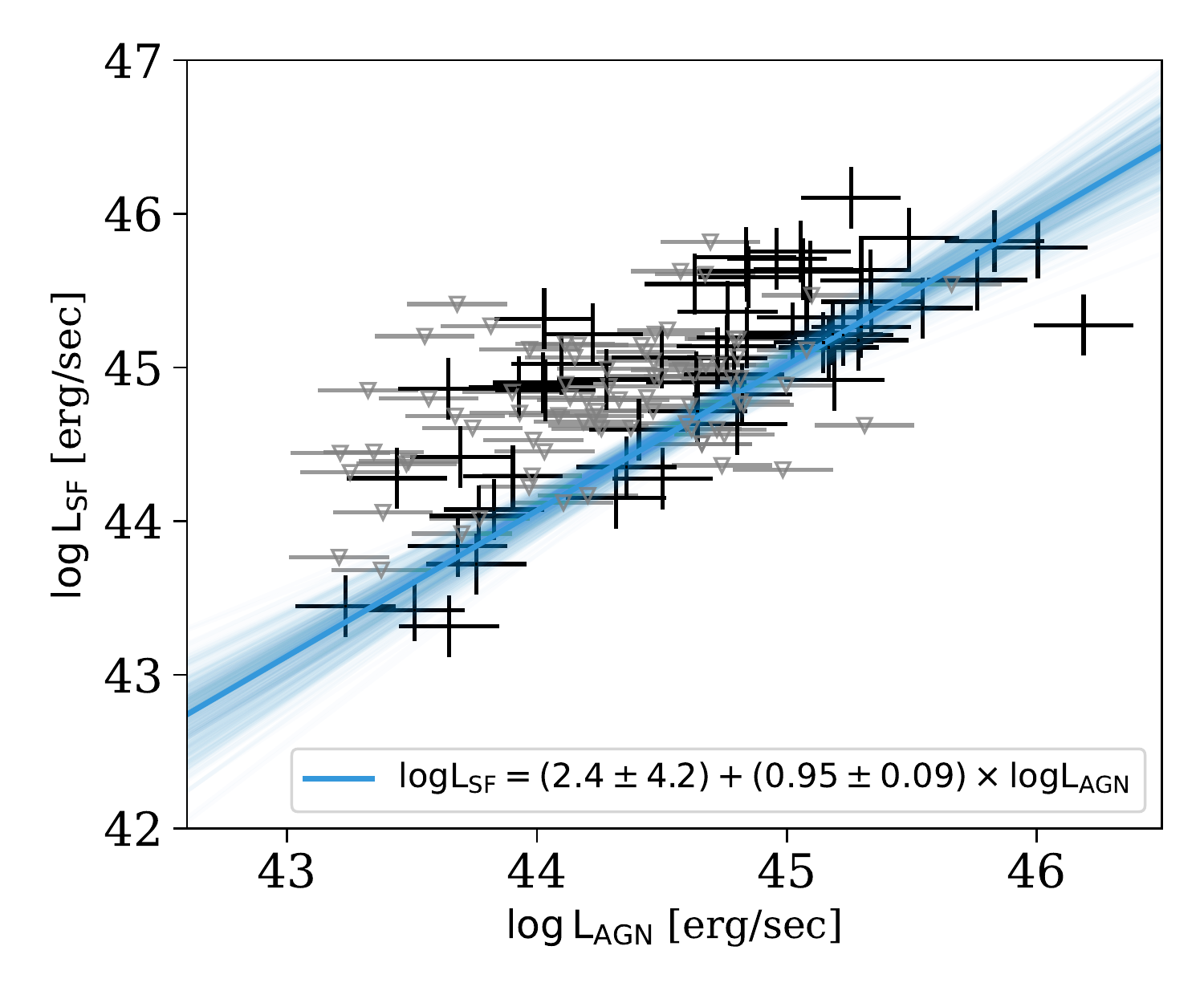}
\caption{\textbf{The SF luminosity versus the AGN luminosity for the objects in our sample.} 
The black markers represent FIR-detected sources and the grey markers represent upper limits. 
The regression was performed with {\sc linmix} and is plotted with a blue line. The fainter blue lines represent the uncertainty of the fit.
 }\label{f:LSF_vs_LAGN}
\end{figure}

In figure \ref{f:LSF_vs_LAGN} we show the relation between $\mathrm{L_{SF}}$ and $\mathrm{L_{AGN}}$. We find a significant correlation between the two. The best-fitting linear relation obtained using {\sc linmix} is $\mathrm{\log L_{SF} = (-4.0 \pm 5.0) + (1.09 \pm 0.11)\log L_{AGN}}$. The slope of the best-fitting relation is steeper than the slope derived by \citet{lani17} for Palomar-Green quasars which are on the SFMS. It is also steeper than the slope found by \citet{netzer09} for a population of type-II AGN from SDSS. The slope is consistent with that of \citet{lani17} for Palomar-Green quasars \emph{above} the SFMS, where both seem to follow roughly a 1:1 relation between $\mathrm{L_{SF}}$ and $\mathrm{L_{AGN}}$.

\begin{figure*}
	\centering
\includegraphics[width=0.95\textwidth]{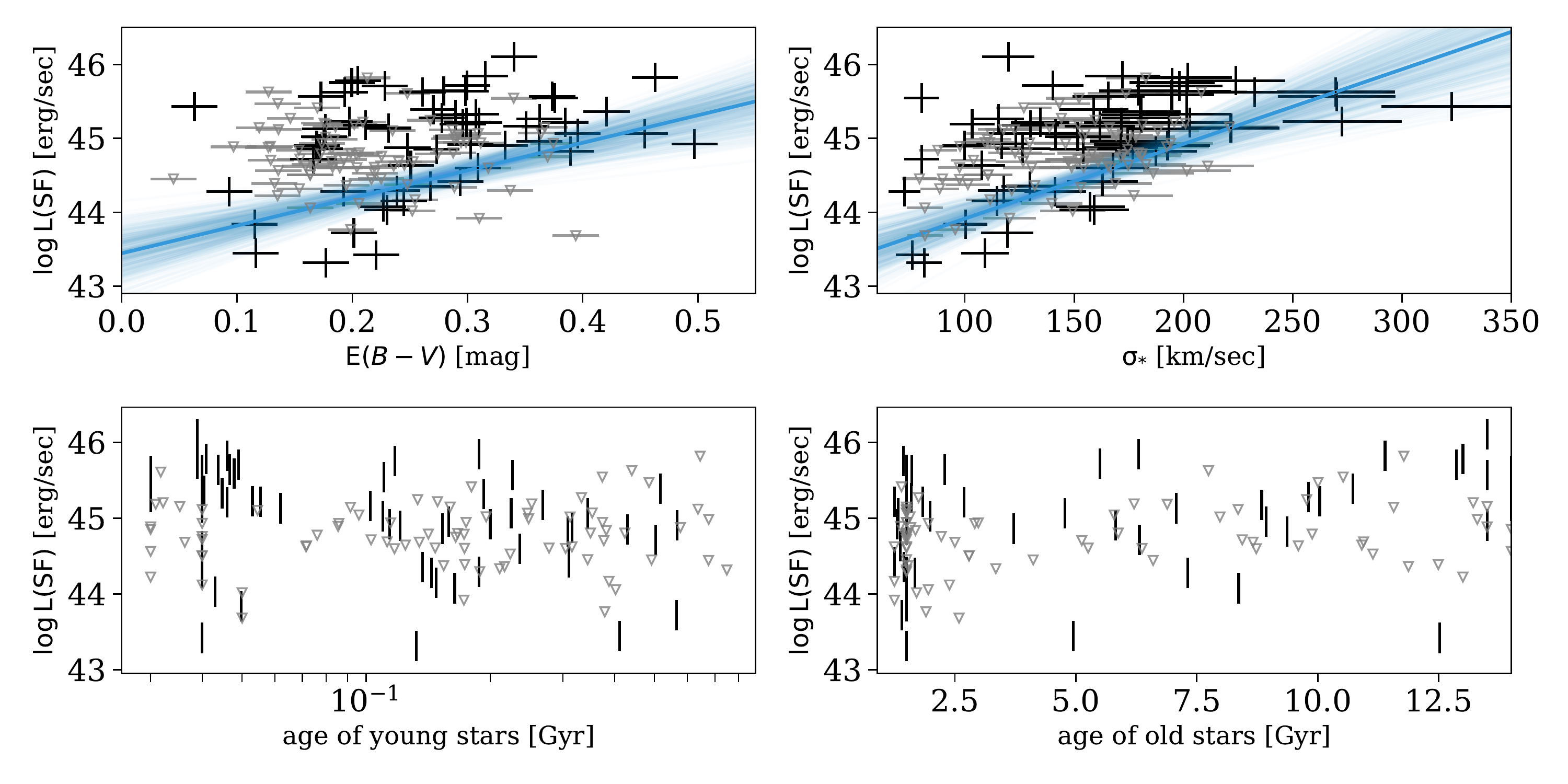}
\caption{\textbf{Relations between the SF luminosity and optical-based stellar properties.} 
The different panels show $\mathrm{L_{SF}}$ versus stellar properties derived from fitting stellar population synthesis models to the optical spectra. The panels show the color excess towards the stars, the stellar velocity dispersion, the mass-weighted age of stars younger than 1 Gyr (young stars), and the mass-weighted age of stars older than 1 Gyr (old stars).
The black markers represent FIR-detected sources and the grey markers upper limits. 
For the statistically-significant correlations at the top, we show the best-fitting lines in blue, and the range of uncertainty in fainter blue.
 }\label{f:LSF_versus_optical_stellar_properties}
\end{figure*}

In figure \ref{f:LSF_versus_optical_stellar_properties} we show the relation between $\mathrm{L_{SF}}$ and various optical-based stellar properties, derived from fitting stellar population synthesis models to the optical spectra of the sources. We examine the relations with the dust reddening towards the stars, the stellar velocity dispersion, the mass-weighted age of stars younger than 1 Gyr (hereafter young stars), and the mass-weighted age of stars older than 1 Gyr (hereafter old stars). Including upper limits in the analysis, we find significant correlations of $\mathrm{L_{SF}}$ with the dust reddening (generalized Kendall's rank correlation coefficient $\tau = 0.19$, p-value=0.0010), and with the stellar velocity dispersion ($\tau = 0.22$, p-value=0.00013). There are no significant correlations between $\mathrm{L_{SF}}$ and the ages of young and old stars. This suggests that the optical spectra do not fully represent the SF nature of these sources. It also suggests that other optically-derived properties, such as the time that passed since the starburst, do not trace the ongoing SF in these objects.

\begin{figure*}
	\centering
\includegraphics[width=0.95\textwidth]{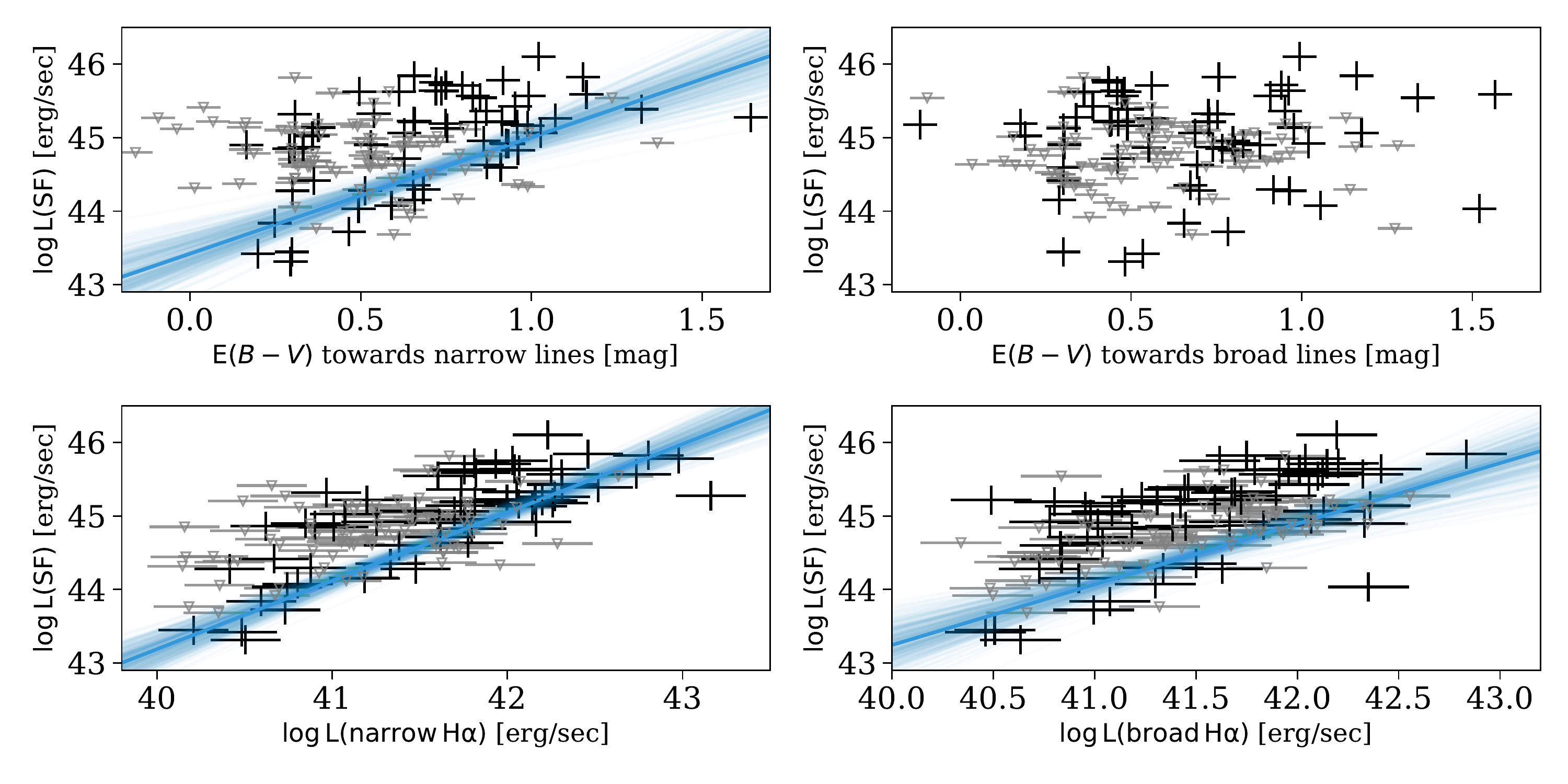}
\caption{\textbf{Relations between the SF luminosity and emission line properties.} 
The top row shows $\mathrm{L_{SF}}$ versus the dust reddening towards the narrow (left panel) and broad emission lines (right panel).
The bottom row shows $\mathrm{L_{SF}}$ versus the dust-corrected narrow (left panel) and broad (right panel) H$\alpha$ luminosity.
The black markers represent FIR-detected sources and the grey markers upper limits. 
The blue lines are as in figure \ref{f:LSF_versus_optical_stellar_properties}.}\label{f:LSF_versus_emis_line_properties}
\end{figure*}

In figure \ref{f:LSF_versus_emis_line_properties} we show the relations between $\mathrm{L_{SF}}$ and several emission line properties. The top row shows the relations with the dust reddening towards the narrow and broad lines. Including the upper limits, we find a significant correlation between $\mathrm{L_{SF}}$ and $\mathrm{E}_{B-V}$ towards the narrow lines, with $\tau=0.25$ and p-value=$1.9\times10^{-5}$. The correlation between $\mathrm{L_{SF}}$ and $\mathrm{E}_{B-V}$ towards the broad lines is not significant. The bottom row of figure \ref{f:LSF_versus_emis_line_properties} shows the relations with the dust-corrected H$\alpha$ luminosity of the narrow and broad components. We find significant correlations in both of the cases, with $\tau = 0.35$ and p-value=$2.8 \times 10^{-9}$ for the narrow H$\alpha$, and $\tau = 0.22$ and p-value=$9.2 \times 10^{-5}$ for the broad H$\alpha$. Since our objects were selected to be dominated by AGN photoionization, we expect their emission line luminosities to be related to the AGN luminosity. Therefore, the correlation between $\mathrm{L_{SF}}$ and the H$\alpha$ luminosity is probably the result of the correlation between $\mathrm{L_{SF}}$ and $\mathrm{L_{AGN}}$.

\begin{figure*}
	\centering
\includegraphics[width=0.7\textwidth]{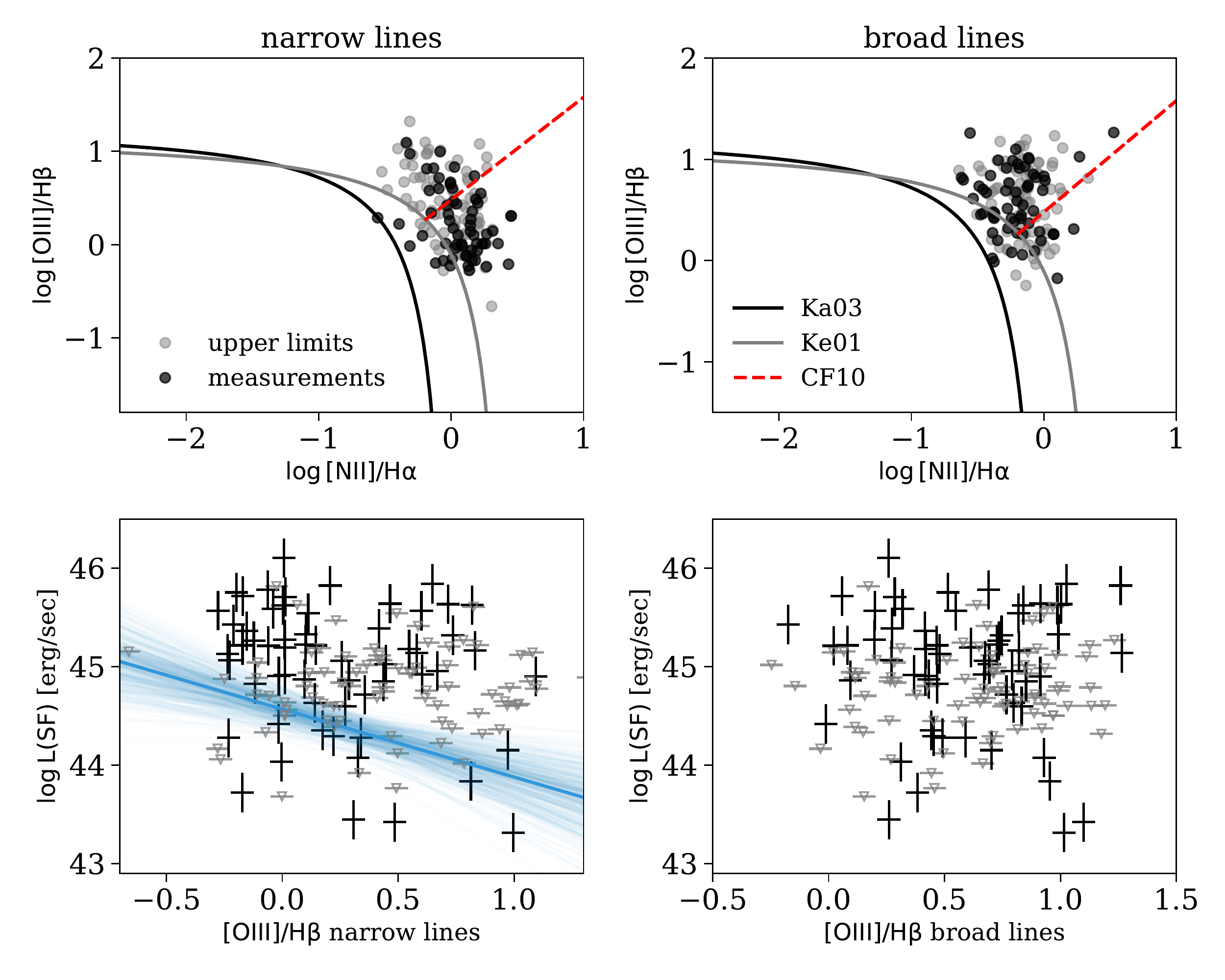}
\caption{\textbf{Relations between the SF luminosity and the ionization state of the gas.} 
\textbf{The top row} shows the location of our sources on the [NII]-based BPT diagram, where we mark FIR-detected sources with black and upper limits with grey. The left panel shows the narrow lines, and the right panel the broad lines. We mark the two separating criteria that are used to separate star forming from AGN-dominated galaxies (\citealt{kewley01, kauff03a}; Ke01 and Ka03 respectively), and the LINER-Seyfert separating line from \citet[CF10]{cidfernandes10}. 
\textbf{The bottom row} shows $\mathrm{L_{SF}}$ versus the narrow and broad [OIII]/H$\beta$ line ratios. The [OIII]/H$\beta$ ratio is closely related to the ionization state of the gas, where log[OIII]/H$\beta \sim 1$ represents Seyferts and log[OIII]/H$\beta \sim 0$ represents LINERs. 
The blue lines are as in figure \ref{f:LSF_versus_optical_stellar_properties}.
 }\label{f:LSF_versus_emis_line_ratio}
\end{figure*}

In figure \ref{f:LSF_versus_emis_line_ratio} we examine the relation between $\mathrm{L_{SF}}$ and the ionization state of the stationary and outflowing gas. The top row shows the location of the systems on the [NII]-based BPT diagram. Interestingly, for the narrow lines, many of the FIR-detected sources are consistent with LINER-like photoionization. To further examine the relation with $\mathrm{L_{SF}}$, in the bottom row of figure \ref{f:LSF_versus_emis_line_ratio} we show $\mathrm{L_{SF}}$ versus the narrow and broad [OIII]/H$\beta$ line ratio. The [OIII]/H$\beta$ line ratio is closely related to the ionization state of the gas and can be used to predict the ionization parameter (see e.g., \citealt{baron19b}). A large ratio of $\mathrm{\log [OIII]/H\beta} \sim 1$ suggests higher ionization parameter, and a small ratio of $\mathrm{\log [OIII]/H\beta} \sim 0$ suggests lower ionization parameter. We find a marginally-significant correlation between $\mathrm{L_{SF}}$ and the narrow [OIII]/H$\beta$ line ratio, with $\tau=-0.12$ and p-value=0.031. It suggests that systems with higher SF luminosity have lower ionization parameter in their stationary gas. This correlation cannot be due to mixing of SF and AGN ionization (e.g., \citealt{wild10, davies14}), since most of the objects in the sample are pure AGN. The correlation can be driven by the column of dust and gas along the line-of-sight to the central source, since these properties correlate with $\mathrm{L_{SF}}$. Since this correlation is only marginally-significant, we do not discuss it any further.

Finally, we discuss the relation of $\mathrm{L_{SF}}$ with the neutral gas properties. To test this, we divide our sample into three groups. The first consists of systems in which we do not detect any NaID (75 systems). The second consists of systems with NaID absorption only (48 systems), and the third with NaID absorption and emission (21 systems). Including upper limits on $\mathrm{L_{SF}}$ in the analysis and using the KME, we find median $\mathrm{L_{SF}}$ values of $\mathrm{\log \Big( L_{SF}/[erg/sec] \Big) = 43.85}$, 44.64, and 44.92 respectively. These values suggest that systems with more luminous SF are more likely to show NaID absorption. They also suggest that systems with a combination of NaID absorption and emission tend to be more luminous in the FIR than systems with absorption alone. We also examined the relation between $\mathrm{L_{SF}}$ and the EW of the NaID absorption, but did not find a significant correlation between the two. However, we do find significant correlations between EW(NaID) and various tracers of the dust reddening (see figure \ref{f:naid_vs_dust} in the appendix). These correlations are known and are the subject of ongoing research (e.g., \citealt{veilleux20, rupke21} and references therein).

\underline{To summarize:} we find a correlation between $\mathrm{L_{SF}}$ and $\mathrm{L_{AGN}}$, with a best-fitting slope that is consistent with that found for quasars above the SFMS. We find correlations between $\mathrm{L_{SF}}$ and various tracers of dust column density ($\mathrm{E}_{B-V}$ towards the stars and towards the narrow lines), suggesting that more luminous SF systems are more dusty. We find a correlation between $\mathrm{L_{SF}}$ and the ionization state of the stationary ionized gas, and suggest that it is related to the dust column density towards the central AGN. Finally, we find a connection between $\mathrm{L_{SF}}$ and the detection of NaID absorption/emission. Systems with detected NaID absorption and emission tend to be the most luminous in the FIR, followed by systems with detected NaID absorption and no emission, and then systems where NaID is not detected at all.

\subsection{Neutral and ionized wind properties}\label{s:results:outflow_properties}

In this section we study the ionized and neutral outflow properties in our sources, and compare them to those derived in other studies. We discuss the general outflow properties (section \ref{s:results:outflow_properties:general}), study the multi-phased nature of the winds (section \ref{s:results:outflow_properties:multiphased}), and check what is the main driver of the observed outflows (section \ref{s:results:outflow_properties:driver}).

\subsubsection{General properties}\label{s:results:outflow_properties:general}

To compare between the ionized outflow properties in post starburst galaxies to those observed in other sources, we use three different samples. The first sample includes local type II AGN from SDSS ($0.05 < z < 0.15$; \citealt{baron19b}). For this sample, the outflowing gas mass, mass outflow rate, and kinetic power, were derived using a combination of 1D spectroscopy from SDSS and multi-wavelength photometry. The second sample is taken from \citet{fiore17}. This is a compilation of outflow properties in active galaxies from the literature, all of which were derived using spatially-resolved spectroscopy. The outflow mass and energetics were derived by applying uniform assumptions about the electron densities and outflow velocities. The sample consists of both local and high-z objects. We restrict ourselves to $z < 0.2$ objects from the \citet{fiore17} sample, which includes the ULIRGs presented in \citet{rupke13} and the AGN presented in \citet{harrison14}. The third sample consists of local AGN ($z < 0.1$; \citealt{bae17}), observed with integral-field spectroscopy.

\begin{table*}
 \caption{\textbf{Comparison of ionized outflow properties in different samples.} The table lists the medians and the (15\%, 85\%) percentiles of the distributions of different outflow properties. }\label{tab:ionized_outflow_comp}
 	\centering
\begin{tabular}{|m{1.5cm} l | c | c | c | c | c| } 
\hline
Sample &  & $\mathrm{v_{max}\,[km/sec]}$ & $\log \Big( n_{\mathrm{e}}/\mathrm{cm^{-3}} \Big)$ & $\log \Big( M_{\mathrm{out}}/\mathrm{M_{\odot}} \Big)$ & $\log \Big( \dot{M}_{\mathrm{out}}/ \mathrm{M_{\odot}\, yr^{-1}} \Big)$ & $\log \Big( \dot{E}_{\mathrm{out}}/\mathrm{erg\, sec^{-1}} \Big)$   \\
\hline
\hline
\citet{fiore17} & & -1024 (-1520, -628) & 2.3 (2.3, 2.3) & - & 1.13 (0.23, 1.64) & 42.6 (41.2, 43.3) \\
\hline
\citet{bae17} & & -800 (-1101, -562) & 2.5 (2.1, 2.8) & 6.1 (5.7, 6.4) & 0.50 (0.03, 0.90) & 41.9 (41.0, 42.4) \\
\hline
\citet{baron19b} & & -823 (-1100, -654) & 4.5 (2.7, 5.1) & 4.3 (3.3, 6.1) & -1.44 (-2.25, -0.44) & 39.3 (38.3, 40.3) \\
\hline
          & r = 0.3 kpc & -766 (-1016, -595) & 4.2 (3.5, 4.9) & 4.8 (4.1, 5.6) & -0.76 (-1.44, 0.09) & 40.5 (39.8, 41.4) \\
This work & r = 1.0 kpc & -766 (-1016, -595) & 3.1 (2.4, 3.9) & 5.8 (5.1, 6.6) & -0.24 (-0.91, 0.61) & 41.0 (40.3, 41.9) \\
          & r = 3.0 kpc & -766 (-1016, -595) & 2.2 (1.5, 2.9) & 6.8 (6.1, 7.6) & 0.23 (-0.44, 1.08) & 41.5 (40.8, 42.4) \\	  
\hline 
\end{tabular}
\end{table*}

\begin{figure*}
	\centering
\includegraphics[width=1\textwidth]{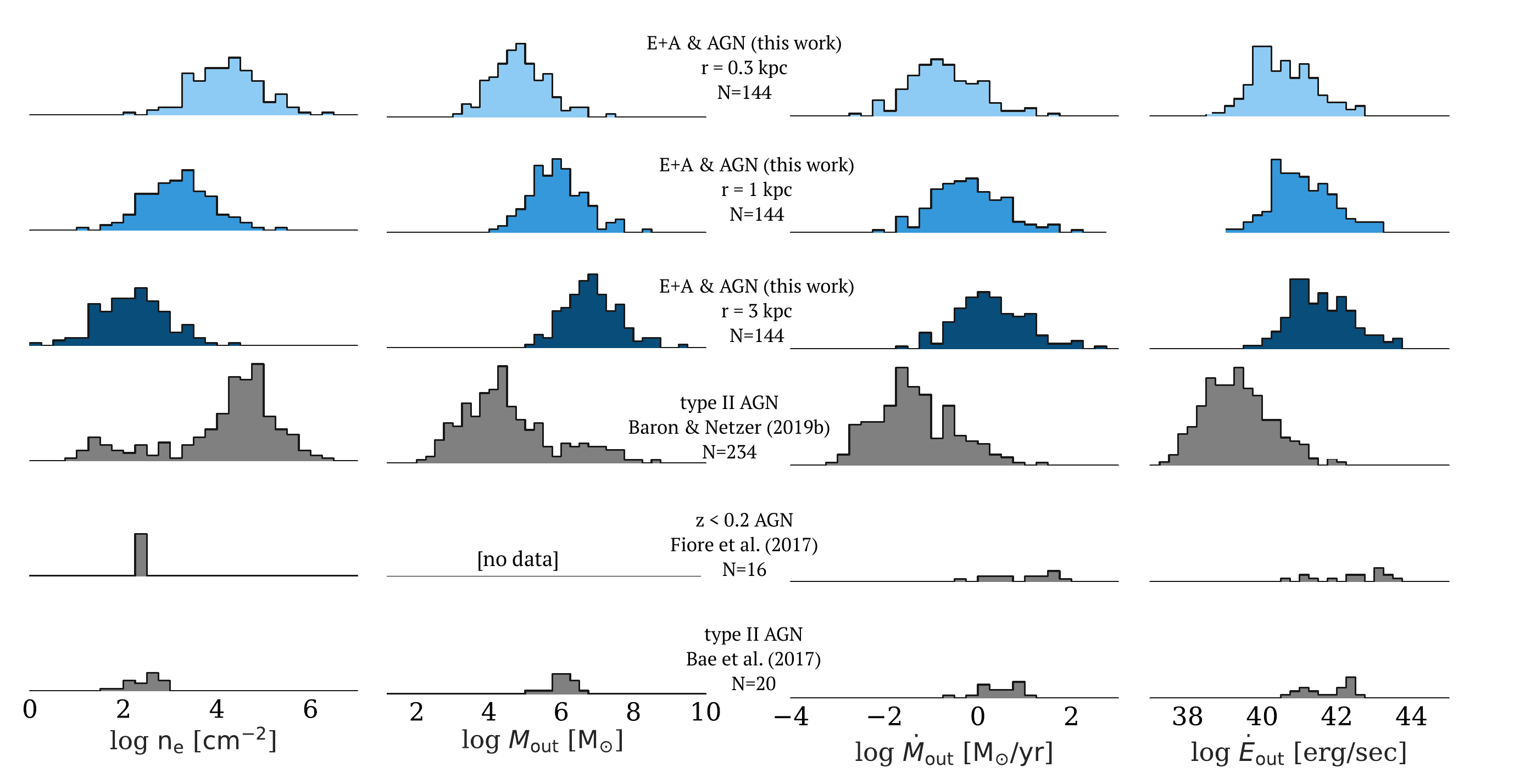}
\caption{\textbf{Comparison of ionized outflow properties in different samples.} 
Each column presents a different property: electron density, outflowing gas mass, mass outflow rate, and kinetic power. 
The outflow properties depend on its extent, and we examine three possibilities for the objects in our sample: 0.3, 1, and 3 kpc.
The uncertainty due to the unknown location exceeds all other uncertainties on the shown properties.
Each row represents the distribution of the outflow property in a different sample, where the samples are indicated in the diagram.
The mass outflow rates and kinetic powers in post starburst galaxies are larger than those observed in typical active galaxies (the \citealt{baron19b} sample), and are more consistent with those observed in systems with high mass outflow rates (the samples by \citealt{fiore17} and \citealt{bae17}).
 }\label{f:ionized_outflow_M_and_E_comparison}
\end{figure*}

While all the three comparison samples consist of local AGN, they differ in their mass outflow rates and kinetic powers. The \citet{fiore17} sample consists of the objects with the highest mass outflow rates, followed by the sample by \citet{bae17}, and then the sample by \citet{baron19b}. This is because the first two samples are based on spatially-resolved spectroscopy, which requires a significant flux in the broad emission lines to reach sufficient signal-to-noise ratio. Such systems have, on average, higher mass outflow rates and kinetic powers. The sample by \citet{baron19b} is not based on spatially-resolved observations and includes systems with lower mass outflow rates and kinetic powers.

\begin{table*}
 \caption{\textbf{Comparison of neutral outflow properties in different samples.} The table lists the medians and the (15\%, 85\%) percentiles of the distributions of different outflow properties. }\label{tab:neutral_outflow_comp}
 	\centering
\begin{tabular}{|l l | c | c | c | c| } 
\hline
Sample &  & $\mathrm{v_{max}\,[km/sec]}$ & $\log \Big( M_{\mathrm{out}}/\mathrm{M_{\odot}} \Big)$ & $\log \Big( \dot{M}_{\mathrm{out}}/ \mathrm{M_{\odot}\, yr^{-1}} \Big)$ & $\log \Big( \dot{E}_{\mathrm{out}}/\mathrm{erg\, sec^{-1}} \Big)$   \\
\hline
\hline
\citet{cazzoli16} & & -294 (-514, -115) & 8.0 (7.8, 8.4) & 1.08 (0.78, 1.41) & 41.8 (41.1, 42.5) \\
\hline
Rupke+ (2005) (U)LIRGs & & -416 (-516, -242) & 9.0 (8.7, 9.3) & 1.5 (1.1, 2.1) & 42.1 (41.4, 42.9) \\ 
\hline
Rupke+ (2005) (U)LIRGs + AGN & & -456 (-694, -371) & 8.7 (8.2, 9.2) & 1.24 (0.89, 1.99) & 42.2 (41.6, 42.9) \\
\hline
          & r = 0.3 kpc & -633 (-855, -476) & 5.8 (5.3, 6.2) & 0.12 (-0.32, 0.61) & 41.1 (40.7, 41.9) \\
This work & r = 1.0 kpc & -633 (-855, -476) & 6.9 (6.4, 7.3) & 0.64 ( 0.20, 1.13) & 41.6 (41.2, 42.4) \\
          & r = 3.0 kpc & -633 (-855, -476) & 7.8 (7.3, 8.2) & 1.12 ( 0.68, 1.61) & 42.1 (41.7, 42.9) \\
\hline 
\end{tabular}
\end{table*}

\begin{figure*}
	\centering
\includegraphics[width=1\textwidth]{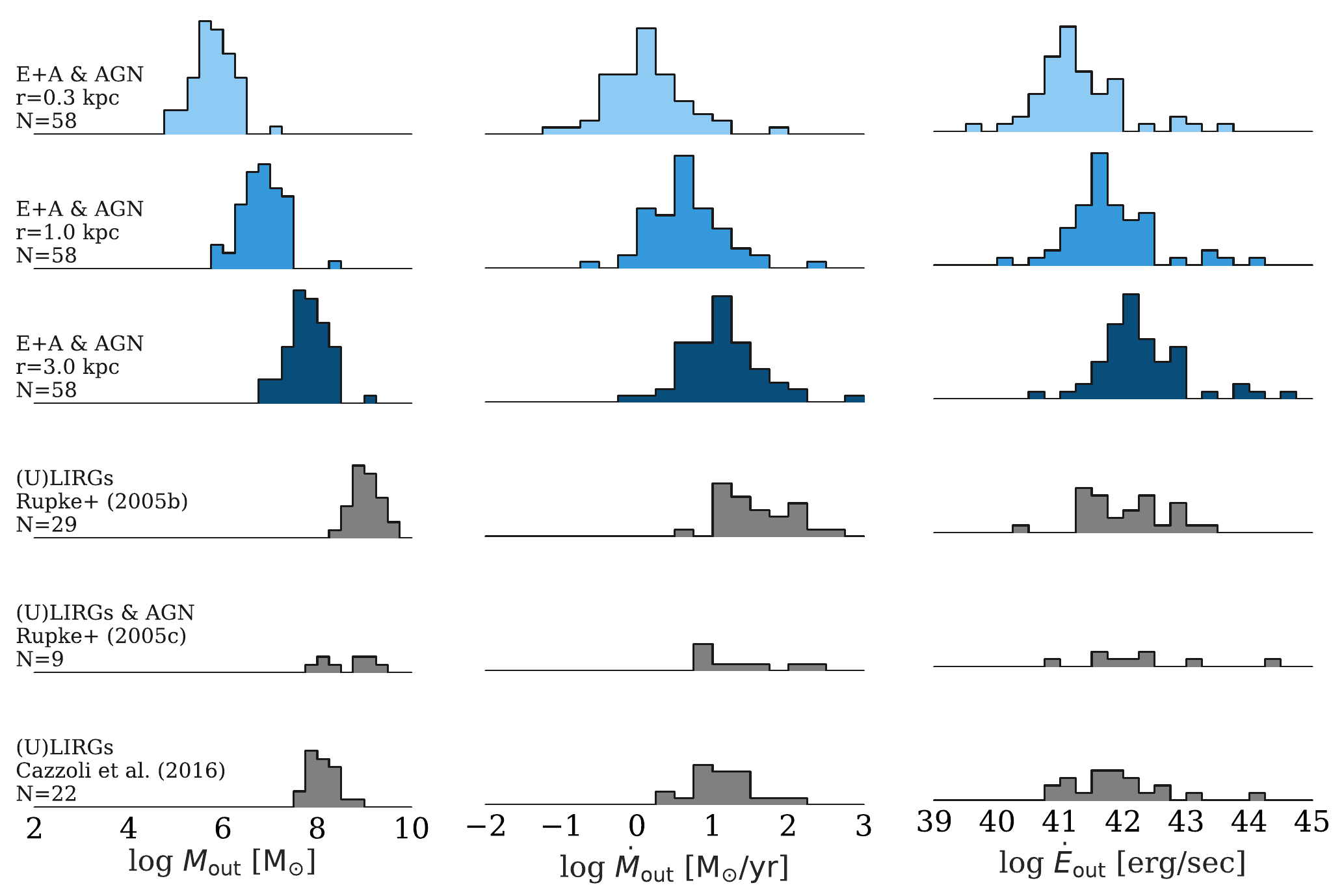}
\caption{\textbf{Comparison of neutral outflow properties in different samples.} 
Each column presents a different outflow property: outflowing gas mass, mass outflow rate, and kinetic power. All the properties depend on the assumed outflow extent, and we examine three possible extents: 0.3, 1, and 3 kpc. The uncertainty due to the unknown location exceeds all other uncertainties on the shown properties. 
The mass, mass outflow rate, and kinetic power in post starburst E+A galaxies are smaller than those observed in (U)LIRGs, for all assumed extents.
 }\label{f:neutral_outflow_M_and_E_comparison_ver2}
\end{figure*}

In table \ref{tab:ionized_outflow_comp} we present summary statistics of the ionized outflow properties in the different samples. In figure \ref{f:ionized_outflow_M_and_E_comparison} we compare the electron densities, gas masses, mass outflow rates, and kinetic powers, of the different samples. All the properties depend on the assumed extent, $r$, and lacking spatially-resolved observations, we examine three possible extents for the objects in our sample: 0.3, 1, and 3 kpc\footnote{The only two cases in which outflows in post starburst galaxies were studied with spatially-resolved spectroscopy reveal larger outflow extents of $r > 1$ kpc \citep{baron18, baron20}, which might suggest larger outflow extents in the rest of the population.}. The electron density and gas mass show a strong dependence on $r$, and are thus highly uncertain. The mass outflow rate and kinetic power show a weaker dependence on $r$, and figure \ref{f:ionized_outflow_M_and_E_comparison} suggests that they are roughly in between the values presented by \citet{baron19b} and those presented by \citet{fiore17} and \citet{bae17}. 

 
Although detected in both star forming and active galaxies, neutral outflows have been typically linked to the SF activity, with more massive outflows detected in systems with significant SF, such as (U)LIRGs (e.g., \citealt{rupke05c, sarzi16, perna17, roberts_borsani19}). We therefore compare the neutral wind properties to those observed in three samples of (U)LIRGs, presented and analyzed by \citet{rupke05a}, \citet{rupke05b}, \citet{rupke05c}, and \citet{cazzoli16}. The sample by \citet{rupke05c} includes (U)LIRGs with AGN, while the other samples do not include AGN. 

In table \ref{tab:neutral_outflow_comp} we present summary statistics of the neutral outflow properties in the different samples. In figure \ref{f:neutral_outflow_M_and_E_comparison_ver2} we compare between the gas masses, mass outflow rates, and kinetic powers in the neutral winds of the different samples. All three properties depend on the extent of the neutral outflow, and are thus uncertain to different levels. The most uncertain property is the outflowing gas mass, which depends linearly on the assumed extent. The mass outflow rate and kinetic power have a weaker dependence on the assumed extent. Figure \ref{f:neutral_outflow_M_and_E_comparison_ver2} shows that the outflowing gas mass in post starburst galaxies is smaller than the mass derived in (U)LIRGs for all the assumed extents. In fact, to obtain a roughly similar distribution of gas mass to those observed in (U)LIRGs, we must assume an outflow extent of $r \sim 30$ kpc, which is highly unlikely. For all assumed extents, the mass outflow rate and kinetic power are also smaller than those derived in (U)LIRGs.

\subsubsection{The multi-phase nature of outflows in post starburst E+A galaxies}\label{s:results:outflow_properties:multiphased}

The nature of multi-phased outflows remains largely unconstrained, with only a handful of studies conducted so far (see reviews by \citealt{cicone18, veilleux20}). Different gas phases (molecular, atomic, and ionized) observed in \emph{different} systems show different outflow velocities, covering factors, and masses (e.g, \citealt{fiore17, cicone18, veilleux20}). It is unclear whether the different outflow phases are connected, and what is the dominant gas phase in the outflow. \citet{rupke05c} and \citet{rupke17} conducted a detailed comparison between the neutral and ionized outflow phases in their sample, finding similarities in some of the systems. In \citet{baron19b} we used dust infrared emission to constrain the neutral fraction in the winds in local type II AGN, and found that the neutral gas mass is a factor of a few larger than the ionized gas mass.


We compared between the outflow velocities, gas masses, mass outflow rates, and kinetic powers, of the ionized and neutral winds in our sample. In this comparison, we assumed a similar outflow extent for the neutral and ionized winds. We did not find a significant correlation between the properties. This suggests that there is no global relation between the neutral and ionized phases in all of the systems, albeit the large uncertainties on some of the derived properties. Nevertheless, a detailed inspection of the emission and absorption profiles reveals similar kinematics for the neutral and ionized outflows in some of the sources. One of the best examples is the post starburst E+A galaxy presented in \citet{baron20}, where our analysis of the spatially-resolved spectra revealed that the two phases are in fact part of the same outflow.

\begin{figure*}
	\centering
\includegraphics[width=0.4\textwidth]{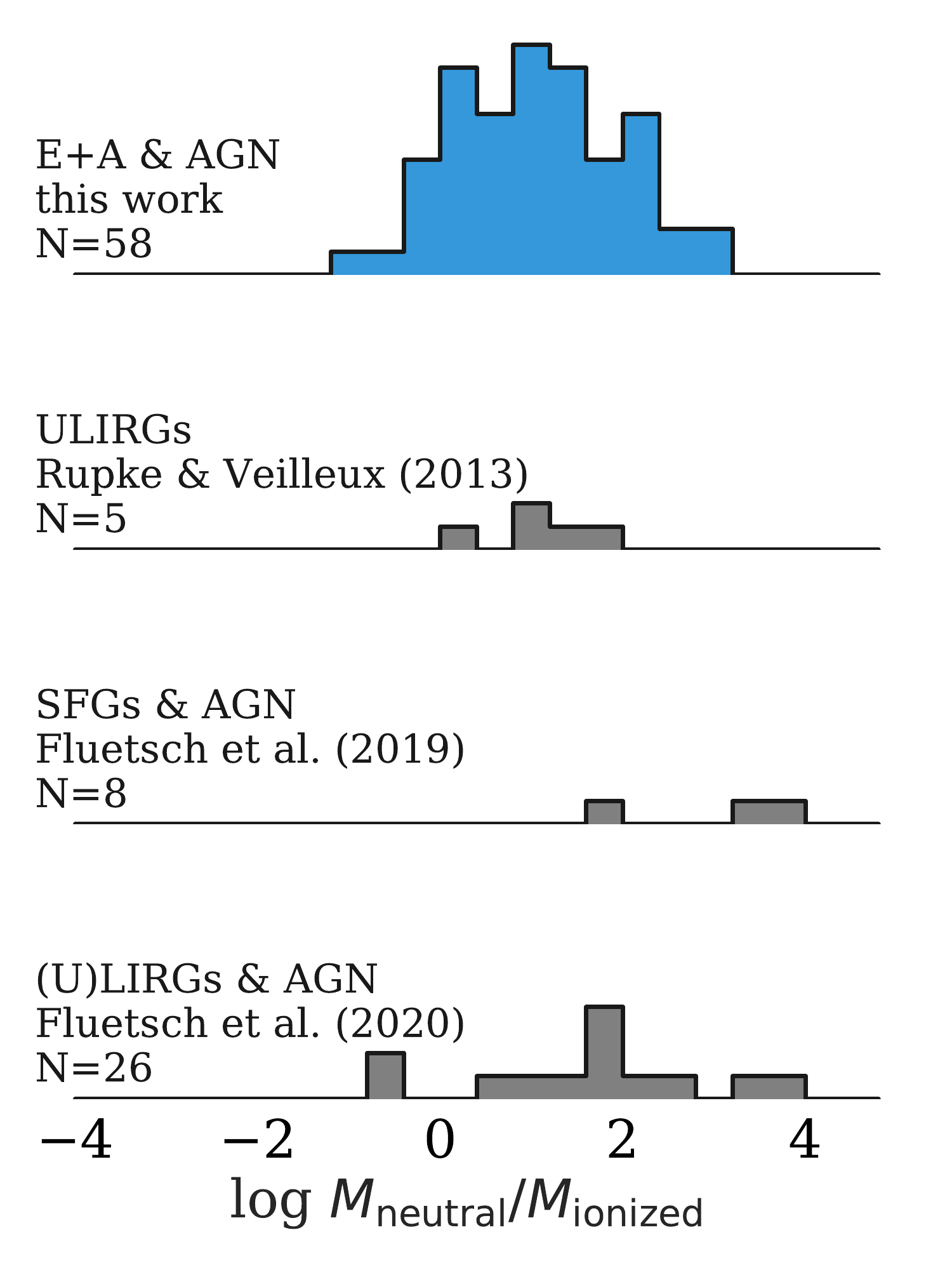}
\includegraphics[width=0.4\textwidth]{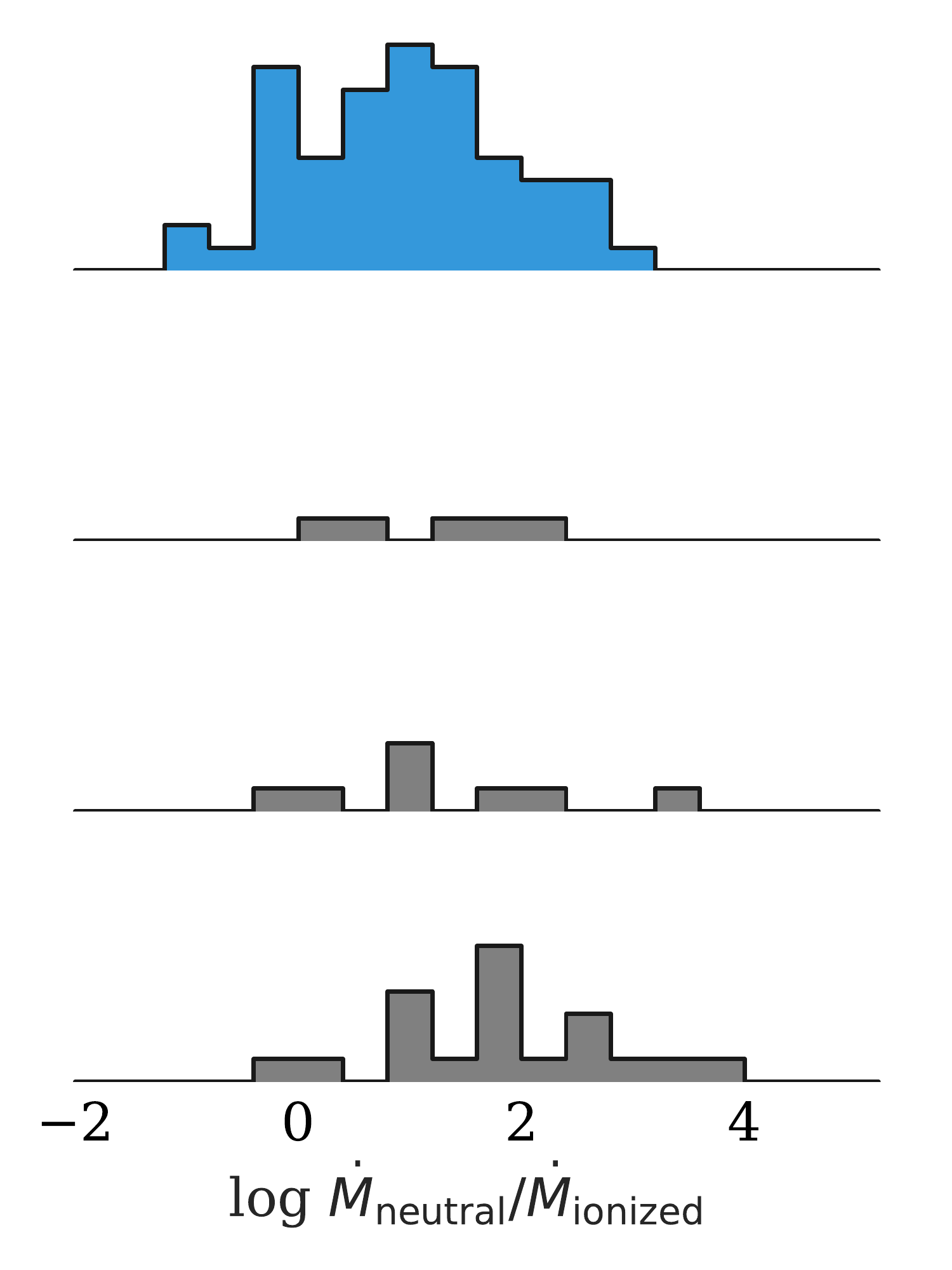}
\caption{\textbf{Neutral-to-ionized mass and mass outflow rate ratios for all samples.} 
The left panels show the neutral-to-ionized gas mass ratio, and the right panels show the mass outflow rate ratio. The samples are marked in the diagram.
The distributions include only sources where both outflow phases were detected. 
 }\label{f:multiphased_outflow_m_and_mdot_comparison}
\end{figure*}

We examine the ratios of mass and mass outflow rate in the ionized and neutral gas and compare them with other samples. The results are shown in figure \ref{f:multiphased_outflow_m_and_mdot_comparison}. We restrict the comparison to sources with detected neutral outflows (58 out of 144 systems). Assuming a similar extent for the neutral and ionized outflows, these ratios do not depend on the assumed wind location. For sources with detected NaID outflows, the figure suggests that the neutral outflow phase is 1--2 orders of magnitude more massive than the ionized outflow phase (in terms of outflowing gas mass and mass outflow rate). However, these distributions do not take into account the large fraction of sources in which we did not detect neutral outflows (86 out of 144), where the ionized outflow phase might be the dominant one.

We compare the $\mathrm{M_{neutral}/M_{ionized}}$ and $\mathrm{\dot{M}_{neutral}/\dot{M}_{ionized}}$ ratios to those derived for three samples from the literature. The first sample consists of ULIRGs \citep{rupke13}, the second consists of SF galaxies with AGN \citep{fluetsch19}, and the third consists of (U)LIRGs with AGN \citep{fluetsch20}. The ratios are roughly consistent with those derived for ULIRGs and are somewhat smaller than those derived by \citet{fluetsch19} and \citet{fluetsch20}. All three samples are small and/or have been collected using heterogeneous selection criteria. In addition, similarly to our sample, upper limits are not reported or included in the analysis.

\subsubsection{What drives the observed winds?}\label{s:results:outflow_properties:driver}

In this section we use correlation analysis to examine what is the main driver of the ionized and neutral winds: AGN activity, SF, or a combination of the two. Since there is a significant correlation between $\mathrm{L_{AGN}}$ and $\mathrm{\mathrm{L_{SF}}}$ (see section \ref{s:results:SF_correlations}), we would expect to find correlations between the outflow properties and both $\mathrm{L_{AGN}}$ and $\mathrm{\mathrm{L_{SF}}}$. To break this degeneracy, we apply partial correlation analysis, as described below.

We start by estimating the correlations between various wind properties ($v_{\mathrm{out}}$, $\dot{M}_{\mathrm{out}}$, $\dot{E}_{\mathrm{out}}$) and $\mathrm{L_{AGN}}$ and $\mathrm{\mathrm{L_{SF}}}$ separately. Here we assumed an outflow extent of $r = 1$ kpc, and to avoid the dependency on $\mathrm{L_{AGN}}$ for the ionized outflow, we estimated $\dot{M}_{\mathrm{out}}$ and $\dot{E}_{\mathrm{out}}$ under the assumption of $n_e = 10^3\,\mathrm{cm^{-3}}$ (see section \ref{s:data_analysis:SDSS:outflow} for additional details). We list the Kendall's rank correlation coefficients, their associated p-values, and the best-fitting parameters of the linear regression in table \ref{tab:lsf_lagn_corrs} in the appendix. The corresponding diagrams are also shown in appendix \ref{a:corr_to_SF_and_AGN}.

Since current Survival Analysis techniques are limited to the case where only the dependent variable has upper limits, we defined the dependent variable to be either $\mathrm{L_{AGN}}$ or $\mathrm{\mathrm{L_{SF}}}$, and the independent variable to be one of the outflow properties. In addition, as noted in section \ref{s:data_analysis:SDSS:outflow}, our adopted uncertainties of the outflow properties reflect only the uncertainty related to the line-fitting process. These uncertainties do not include the unknown wind location and geometry. Therefore, the values listed in table \ref{tab:lsf_lagn_corrs} should not be compared to equivalent correlations reported by other studies, where $\mathrm{L_{AGN}}$ and  $\mathrm{\mathrm{L_{SF}}}$ were defined as the independent variables. We use these only to compare between the relative importance of the AGN and the SF in driving the winds, through a partial correlation analysis. 

The partial correlation measures the correlation between two properties, (x, y$_{1}$), with the effect of a different controlling property, y$_{2}$, removed. In the present case y$_{1}$ and y$_{2}$ are $\mathrm{L_{AGN}}$ or $\mathrm{\mathrm{L_{SF}}}$. For each pair of variables (e.g., $\mathrm{v_{out}}$ versus $\mathrm{L_{AGN}}$), we estimated the partial correlation coefficient between the variables, while controlling for the third dependent variable ($\mathrm{L_{SF}}$ in this case). We calculated the partial correlation coefficients using the linear regression results for the three different pairs, e.g., ($\mathrm{v_{out}}$, $\mathrm{L_{AGN}}$), ($\mathrm{v_{out}}$, $\mathrm{L_{SF}}$), and ($\mathrm{L_{SF}}$, $\mathrm{L_{AGN}}$) for the example above. Since the correlations between the neutral outflow properties and $\mathrm{L_{AGN}}$ and $\mathrm{\mathrm{L_{SF}}}$ were not significant, we restricted the partial correlation analysis of the ionized outflow properties.

In table \ref{tab:lsf_lagn_ionized_partial_corr} we list the partial correlation coefficients and their associated p-values for the ionized outflow properties versus $\mathrm{L_{AGN}}$ and $\mathrm{\mathrm{L_{SF}}}$. For the maximum outflow velocity, the partial correlation with $\mathrm{L_{AGN}}$ is more significant than that with $\mathrm{\mathrm{L_{SF}}}$. This might suggest that the AGN plays a more important role in accelerating the ionized gas. For the mass outflow rate and kinetic power, we do not find notable differences between $\mathrm{L_{AGN}}$ and $\mathrm{\mathrm{L_{SF}}}$ in terms of the partial correlation coefficients or their p-values. Therefore, we are not able to identify the main driver of the ionized winds.

\begin{table}
 \caption{\textbf{Partial correlations between the ionized outflow properties and $\mathrm{\mathrm{L_{AGN}}}$ and $\mathrm{\mathrm{L_{SF}}}$.}}\label{tab:lsf_lagn_ionized_partial_corr}
 	\centering
\begin{tabular}{|l l | c c|}
\hline
   &   &\multicolumn{2}{c|}{Kendall's partial rank correlation} \\
 x & y & $\tau$ & p-value \\	
\hline
$\mathrm{v_{out}}$ & $\mathrm{\log{} L_{AGN}}$ & 0.18 & 0.0014 \\
$\mathrm{v_{out}}$ & $\mathrm{\log{} L_{SF}}$  & 0.11 & 0.042 \\
\hline

$\mathrm{\log{} \dot{M}_{out}}$ & $\mathrm{\log{} L_{AGN}}$ & 0.18 & 0.0015 \\
$\mathrm{\log{} \dot{M}_{out}}$ & $\mathrm{\log{} L_{SF}}$  & 0.17 & 0.0023 \\
\hline

$\mathrm{\log{} \dot{E}_{out}}$ & $\mathrm{\log{} L_{AGN}}$ & 0.18 & 0.0014 \\
$\mathrm{\log{} \dot{E}_{out}}$ & $\mathrm{\log{} L_{SF}}$  & 0.17 & 0.0024 \\
\hline

\end{tabular}
\end{table}

\section{Summary and Conclusions}\label{s:conclusions} 

E+A galaxies are thought to be the missing link between ULIRGs and red and dead ellipticals, possibly due to AGN-feedback that quenches star formation and stops the growth of their stellar mass. Unfortunately, little is known about AGN feedback in this phase of galaxy evolution.

In this work we constructed the first large sample of post starburst E+A galaxy candidates with AGN and ionized outflows. For this sample of 144 objects, we collected optical spectra from SDSS and FIR photometry from IRAS. We used the FIR observations to study the obscured star formation in these systems and modeled the stellar population and the star formation history. Using the optical spectra, we decomposed the observed emission lines into narrow and broad kinematic components, which correspond to stationary NLR and outflowing gas respectively. On top of the ionized outflows seen in all 144 sources, we also discovered NaID absorption in 69 systems, 21 of which show evidence of NaID \textbf{emission} as well. We modeled the NaID profiles and concluded that 58 out of 69 systems host neutral gas outflows, while the other 11 sources host stationary neutral gas. We used these data to calculate the outflowing gas mass, mass outflow rate, and kinetic power of the wind. Our main results pertain to the SF and outflow properties in post starburst E+A galaxies and are summarized below:

\begin{enumerate}

\item The fraction of post starburst E+A galaxy candidates with emission lines that are above the star forming main sequence ranges from 13\% to 45\%. In addition, the fraction of starbursts ranges from 7\% to 32\%. These suggest that many systems selected to have post starburst signatures based on their optical properties are in fact obscured starbursts (consistent with \citealt{poggianti00}). Optical-based derived properties, such as H$\alpha$-based SFR, Dn4000\AA\, index, and the age of the stellar population, do not represent the true star forming nature of these systems.
\item Post starburst candidates with more luminous emission lines are more likely to be found above the main sequence, with a larger fraction of starbursts.
\item Among AGN hosts, post starburst E+A candidates are more likely to be found above the main sequence compared to typical Seyfert galaxies, with a significantly larger fraction of obscured starbursts (21\% compared to 6\%). 
\item We found a significant correlation between $\mathrm{L_{SF}}$ and $\mathrm{L_{AGN}}$, with a best-fitting slope that is consistent with that derived for quasars above the main sequence.
\item We found correlations between $\mathrm{L_{SF}}$ and various tracers of the dust column density, suggesting that more luminous star forming systems are more dusty.
\item We found a marginally-significant correlation between $\mathrm{L_{SF}}$ and the ionization state of the stationary gas, where systems with detected FIR emission are more likely to be classified as LINERs than as Seyferts.
\item We found a connection between $\mathrm{L_{SF}}$ and the detection of NaID absorption/emission, where systems with detected NaID absorption and emission tend to be the most luminous in the FIR. Following them are systems with NaID absorption only, and then systems where NaID is not detected.

\item The velocities of the ionized and neutral winds are roughly similar to those observed in local AGN and (U)LIRGs.
\item For the ionized outflow, we found that the mass outflow rate and kinetic power of the winds in E+A galaxies are larger than those observed in local active galaxies of similar AGN luminosity, and are consistent with those observed in the tails of the respective distributions.
\item For the neutral outflow, the mass, mass outflow rate, and kinetic power are smaller than those observed in (U)LIRGs with and without AGN.
\item While we found similar kinematics  for the ionized and neutral outflows in some of the sources, there is no global correlation between the outflow velocity, gas mass, mass outflow rate, and kinetic power, in the neutral and ionized winds.
\item Among the sources with detected NaID absorption, the mass and mass outflow rate of the neutral gas are larger than those of the ionized gas by a factor of 10-100. For the 86 out of 144 sources in which NaID is not detected, the dominant outflow phase is not known.
\item Using a partial correlation analysis, we found that the velocity of the ionized outflow seems to be related to $\mathrm{L_{AGN}}$ more than to $\mathrm{L_{SF}}$. As for the mass outflow rate and kinetic power, we were unable to isolate the dominant driver of the winds.
\end{enumerate}

The results presented in this work are based on publicly-available optical 1D spectroscopy and FIR observations. We followed-up a subset of these sources with an optical IFU and mm observations. In \citet{baron18} and \citet{baron20} we presented an analysis of the spatially-resolved optical spectra for two post starburst galaxies. The analysis of the optical IFU data and mm observations of the rest of the objects will be presented in forthcoming publications.

\section{Data availability}\label{s:data_avail}

We make the various estimated properties of the post starburst galaxy candidates in our sample publicly-available. These include the emission and absorption line properties, and the stellar properties, derived from the optical spectra. These also include the derived mass and energetics of the ionized and neutral outflows in our sources. We also make the stellar masses, FIR-based SFRs, and AGN luminosities publicly-available. These data will be made publicly-available through the online supplementary material and Vizier upon the acceptance of the manuscript.

\section*{Acknowledgments}
We thank the referee, H. Yesuf, for useful comments and suggestions that helped improve this manuscript.
We thank O. Almaini and I. Smail for useful comments about our manuscript.
D. Baron is supported by the Adams Fellowship Program of the Israel Academy of Sciences and Humanities.
This research made use of {\sc Astropy}\footnote{http://www.astropy.org}, a community-developed core Python package for Astronomy \citep{astropy2013, astropy2018}.

This work made use of SDSS-III\footnote{www.sdss3.org} data. Funding for SDSS-III has been provided by the Alfred P. Sloan Foundation, the Participating Institutions, the National Science Foundation, and the U.S. Department of Energy Office of Science. SDSS-III is managed by the Astrophysical Research Consortium for the Participating Institutions of the SDSS-III Collaboration including the University of Arizona, the Brazilian Participation Group, Brookhaven National Laboratory, Carnegie Mellon University, University of Florida, the French Participation Group, the German Participation Group, Harvard University, the Instituto de Astrofisica de Canarias, the Michigan State/Notre Dame/JINA Participation Group, Johns Hopkins University, Lawrence Berkeley National Laboratory, Max Planck Institute for Astrophysics, Max Planck Institute for Extraterrestrial Physics, New Mexico State University, New York University, Ohio State University, Pennsylvania State University, University of Portsmouth, Princeton University, the Spanish Participation Group, University of Tokyo, University of Utah, Vanderbilt University, University of Virginia, University of Washington, and Yale University. 

\bibliographystyle{mn2e}
\bibliography{ref_ea_gals_sample}

\clearpage

\onecolumn

\appendix

\section{Comparison between different post starburst selection methods}\label{a:selection}

In this section we compare between the objects selected using our method (section \ref{s:sample_selection}) and those selected using the methods presented in \citet{alatalo16a} and \citet{yesuf17b}. \citet{yesuf17b} selected galaxies with EW(H$\delta$)$> 5$ \AA, using the H$\delta_{F}$ index from SDSS value-added catalog. To select AGN-dominated systems, they used the SDSS value-added catalog and selected objects with emission line ratios consistent with Seyfert-like ionization. We found that 7\% of their objects are not classified as post starbursts by our algorithm, since our estimate of the H$\delta$ EW is smaller than 5 \AA. A manual inspection of the false positives reveals that many of them are objects with low SNR that were mistakenly classified as H$\delta$-strong galaxies by \citet{yesuf17b}. Furthermore, 39 out of our 144 windy post starburst galaxies are not present in the \citet{yesuf17b} catalog, although only 15 are classified as LINERs according to their narrow emission lines. \citet{alatalo16a} selected galaxies with EW(H$\delta$)$> 5$ \AA, using the H$\delta_{A}$ index from SDSS value-added catalog. We found that $\sim$30\% of their objects are not classified as post starbursts by our algorithm, since our estimate of the H$\delta$ EW is smaller then 5 \AA. As noted in the main text, we found that using the H$\delta_{A}$ index results in a large number of false positives. 

\section{Details on the NaID profile fitting}\label{a:naid_fitting}

We modeled the NaID absorption and emission profiles as follows. For the case of a single kinematic component in absorption, we followed \citet{rupke05a} and modeled the profile as:
\begin{equation}\label{eq:2}
	{I(\lambda) = 1 - C_{f} + C_{f}\exp \{-\tau_{K}(\lambda)-\tau_{H}(\lambda) \}},
\end{equation}
where $I(\lambda)$ is the normalized continuum flux, $C_f$ is the gas covering fraction, and $\tau_{K}(\lambda)$ and $\tau_{H}(\lambda)$ are the optical depths of the NaID $K$ ($\lambda$ 5897 \AA) and $H$ ($\lambda$ 5890 \AA) lines respectively. We assumed a Gaussian optical depth for each of the lines:
\begin{equation}\label{eq:3}
	{\tau(\lambda) = \tau_0 e^{-(\lambda - \lambda_0)^2/(\lambda_0 b/c)^2}},
\end{equation}
where $\tau_0$ is the optical depth at line center, $\lambda_0$ is the central wavelength, $b$ is the Doppler parameter ($b = \sqrt{2} \sigma$), and $c$ is the speed of light. In our fits, we set the Doppler parameter $b$ to be the same for the two doublet lines, and tied the central wavelengths to have the same velocity with respect to systemic. The doublet optical depth ratio, $\tau_{0, H}/\tau_{0, K}$, is estimated directly from the curve-of-growth using $\tau_{0, K}$ \citep{draine11}. Therefore, this model has four free parameters: $C_f$, $\lambda_{0, K}$, $b$, and $\tau_{0, K}$. 

\begin{figure*}
	\centering
\includegraphics[width=1\textwidth]{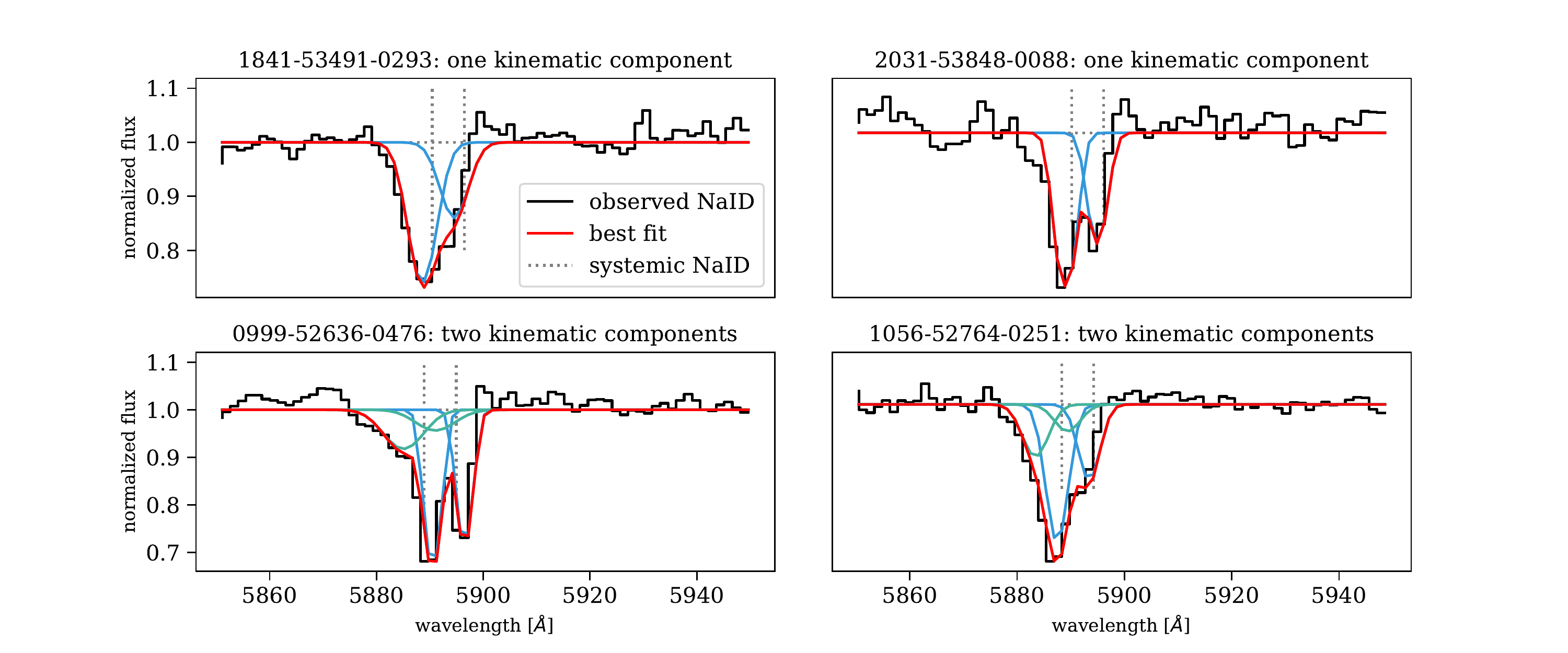}
\caption{\textbf{Four examples of the best-fitting NaID absorption profiles.} 
\textbf{Top:} two examples of cases with a single kinematic component. The black line represents the normalized spectrum, the blue line shows the NaID $K$ and $H$ absorption lines, and the red line represents the best-fitting profile.
\textbf{Bottom:} two examples of cases with two kinematic components. The blue lines represent the first kinematic component and the green lines represent the second. The SDSS identifiers (plate-MJD-fiber) of the galaxies are listed in the title of each panel.
 }\label{f:NaID_absorption_example_fits}
\end{figure*}

For systems that show two kinematic components in absorption, we assumed the simplest case of completely overlapping atoms (see \citealt{rupke05a}) and modeled the profile as:
\begin{equation}\label{eq:4}
	\begin{split}
	& I(\lambda) = 1 - C_{f} + C_{f}\exp \{-\tau_{1, K}(\lambda)-\tau_{1, H}(\lambda) \\
	& -\tau_{2, K}(\lambda)-\tau_{2, H}(\lambda) \},
	\end{split}
\end{equation}
where the subscript $1$ represents the first kinematic component (usually narrower) and $2$ represents the second kinematic component (usually broader and blueshifted). Here too we assumed a Gaussian optical depth. This model has seven free parameters: $C_{f}$, $\lambda_{0, K, 1}$, $b_1$, and $\tau_{0, K, 1}$ for the first component, and $\lambda_{0, K, 2}$, $b_2$, and $\tau_{0, K, 2}$ for the second component. In figure \ref{f:NaID_absorption_example_fits} we show four examples of the best-fitting profiles obtained for the cases of single and two kinematic components. 

\begin{figure*}
	\centering
\includegraphics[width=1\textwidth]{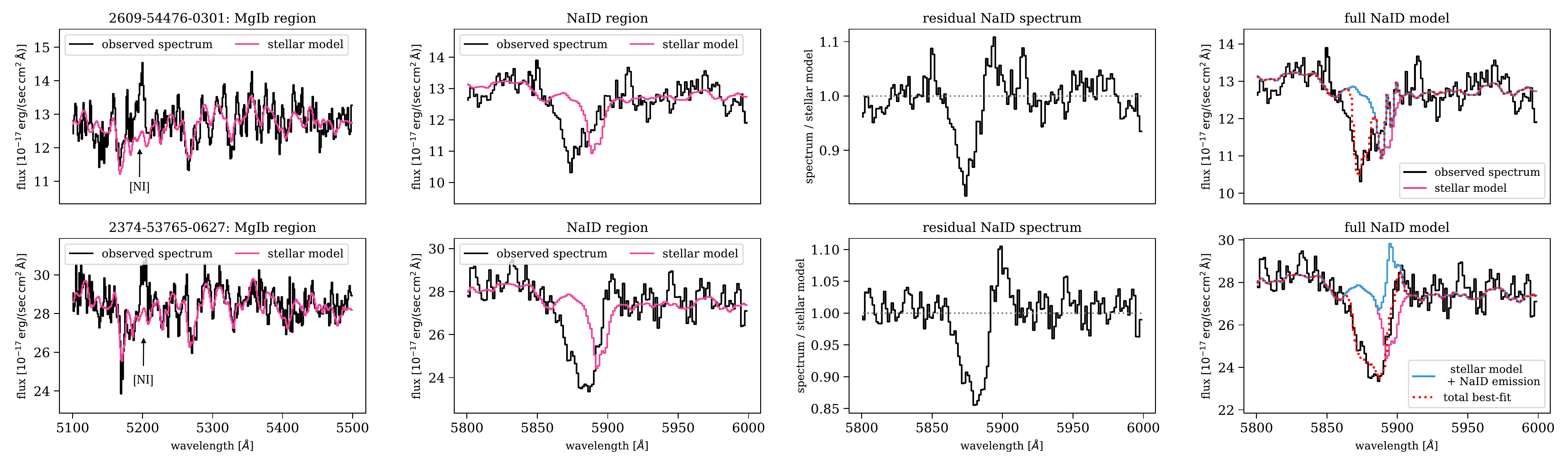}
\caption{\textbf{Examples of best-fitting NaID absorption and emission line profiles.} 
Each row represents a different galaxy, where we list its SDSS identifier (plate-MJD-fiber) in the title of the left-most panel. The first columns show the observed spectra (black) and the best-fitting stellar models (pink) around the MgIb absorption complex. The second columns show the observed spectra and the best-fitting stellar models around the NaID lines. The third columns show the normalized spectra around the NaID lines, where a redshifted emission component is clearly visible. The last columns show our best-fitting models of the spectra, where the pink line shows the stellar model, the blue line shows the stellar model with the NaID emission, and the red dotted line shows the total profile, which is a combination of the stars, the NaID emission, and NaID absorption.
 }\label{f:NaID_absorption_and_emission_example_fits}
\end{figure*}

For systems that show a combination of NaID absorption and emission, we used a slightly simplified version of the model presented in \citet{baron20}. We modeled the observed flux as:
\begin{equation}\label{eq:5}
	\begin{split}
	& f(\lambda) = \Big[f_{\mathrm{stars}}(\lambda) + f_{\mathrm{HeI}}(\lambda) + f_{\mathrm{NaID\, emis}}(\lambda) \Big] \times I_{\mathrm{NaID\, abs}}(\lambda),
	\end{split}
\end{equation}
where $f(\lambda)$ is the observed flux, $f_{\mathrm{stars}}(\lambda)$ is the stellar continuum, $f_{\mathrm{HeI}}(\lambda)$ represents the HeI emission ($\lambda$ 5876 \AA), $f_{\mathrm{NaID\, emis}}(\lambda)$ represents the redshifted NaID emission, and $I_{\mathrm{NaID\, abs}}(\lambda)$ represents the NaID absorption. Since the NaID emission is additive, while the NaID absorption is multiplicative, one must model the observed spectrum rather than the normalized one. Following \citet{baron20}, we made several assumptions to reduce the number of free parameters. We used the best-fitting stellar population synthesis model as the stellar model $f_{\mathrm{stars}}(\lambda)$. The HeI emission is constructed entirely from the best-fitting H$\alpha$ emission profile, assuming the same velocity and velocity dispersion. The HeI to H$\alpha$ intensity ratio depends on the ionization parameter and is taken to be 0.033 \citep{baron20}. Therefore, the stellar and HeI models have no free parameters. We modeled the NaID emission as a sum of two Gaussians:
\begin{equation}\label{eq:6}
	\begin{split}
	& f_{\mathrm{NaID\,emis}}(\lambda) = A_{K} e^{-(\lambda - \tilde{\lambda}_{0,K})^2 / 2 \sigma^2} + A_{H} e^{-(\lambda - \tilde{\lambda}_{0,H})^2 / 2 \sigma^2},
	\end{split}
\end{equation}
where $A_{K}$ and $A_{H}$ are the Gaussian amplitudes, $\tilde{\lambda}_{0,K}$ and $\tilde{\lambda}_{0,H}$ are the central wavelengths, which are tied so that the lines have the same velocity with respect to systemic, and $\sigma$ is the velocity dispersion. The amplitude ratio $A_{H}/A_{K}$ is restricted to the range 1--2, where the limits correspond to the optically-thick and optically-thin regimes respectively. This model has four free parameters: $A_{K}$, $A_{H}$, $\tilde{\lambda}_{0,K}$, and $\sigma$. Finally, we modeled the NaID absorption $I_{\mathrm{NaID\, abs}}(\lambda)$ using the expression for a single kinematic component from equation \ref{eq:2}, which has four free parameters. Therefore, the full emission and absorption model has eight free parameters. In figure \ref{f:NaID_absorption_and_emission_example_fits} we show two examples of best-fitting profiles of this type. 

There are possible degeneracies between the free parameters. In the case of NaID absorption, the strength of the absorption depends on the covering fraction $C_f$ and on the optical depth $\tau$. In principle, one could model the observed strength with a small covering fraction and a large optical depth, or with a large covering fraction and a small optical depth. However, the observed doublet ratio breaks the degeneracy between these parameters. This is because a small optical depth results in a doublet ratio of $\sim$2 (optically-thin regime), while a large optical depth results in a ratio of $\sim$1 (optically-thick regime). In addition, the absorption profile can be modeled with multiple ($> 2$) kinematic components with small optical depths. Our data do not allow us to distinguish between cases of more than one kinematic component. We therefore focus on integrated properties which do not depend on the number of assumed components, such as the total line EW and the maximum velocity of the NaI gas. For the case of NaID emission and absorption, the emission can "fill-in" the absorption, resulting in a degeneracy between the absorption optical depth $\tau$ and the emission amplitude $A$. However, since the NaID emission is redshifted with respect to the absorption, we expect this degeneracy to be minimal. Nevertheless, the best-fitting NaID emission profiles presented here are somewhat uncertain.

\section{Details of other galaxy samples}\label{a:other_samples}

We made use of four different E+A galaxy samples. The first is the sample presented in this study, of E+A galaxies with AGN and ionized outflows. The three other samples were taken from \citet{alatalo16a}, \citet{yesuf17b}, and \citet{french18}. All three studies selected E+A galaxies according to their H$\delta$ absorption strength (see additional details in the papers), but used different cuts on the emission lines. \citet{yesuf17b} selected E+A galaxies with emission line ratios consistent with pure AGN photoionization. Their sample includes only  Seyferts, with no LINERs or composite galaxies. \citet{alatalo16a} selected E+A galaxies with emission line ratios that are consistent with those predicted by shock models. In practice however, this selection includes emission line galaxies with line ratios consistent with photoionization by SF or AGN. \citet{french18} selected quenched E+A galaxies by requiring EW(H$\alpha$ emission) $< 3$ \AA. The redshift distributions of the galaxies in the different samples are similar ($z \sim 0.1$). 

We constructed an additional sample of type II AGN from SDSS. We used the seventh data release of SDSS and the MPA-JHU value-added catalog \citep{b04, kauff03b, t04}. We selected all sources at $0.05 < z < 0.15$ with reliable stellar mass estimates from the MPA-JHU catalog. We selected sources with emission lines measured with signal-to-noise ratios larger than 3 in the following lines: H$\beta$, [OIII], [OI], [NII], and H$\alpha$. Out of these, we selected sources which are classified as Seyferts (no composites or LINERs) according to the [NII]/H$\alpha$ and [OI]/H$\alpha$ diagnostic diagrams. We further selected sources with reliable W4-band magnitudes from WISE \citep{wright10} and available IRAS scans at 60 $\mathrm{\mu m}$. 

\section{Correlations of EW(NaID) with dust reddening}\label{a:corr_naid_dust}

\begin{figure*}
	\centering
\includegraphics[width=1\textwidth]{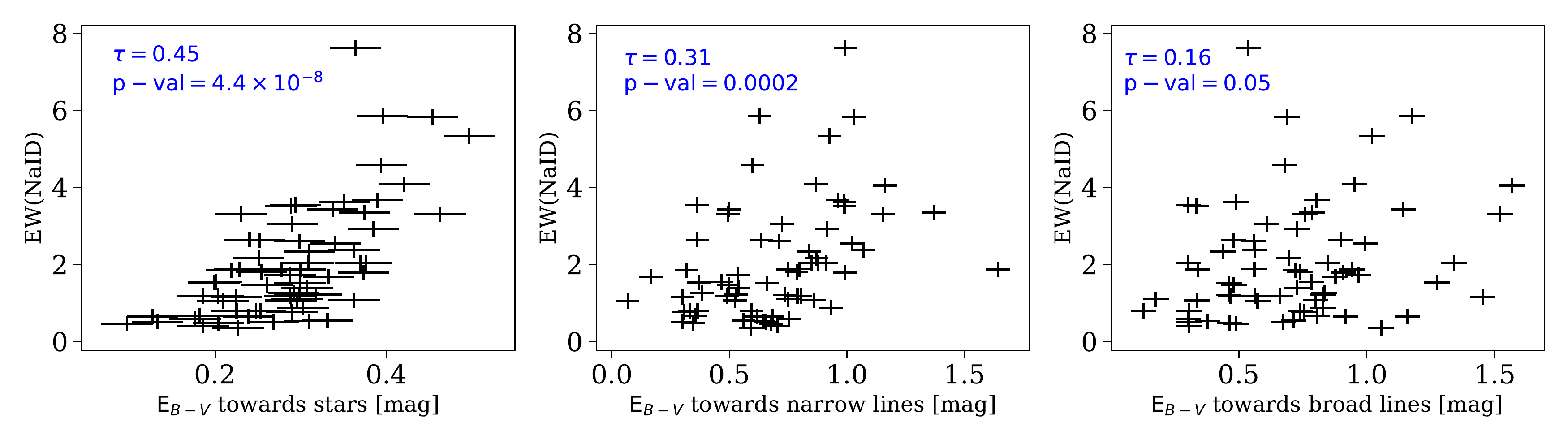}
\caption{\textbf{Correlations of EW(NaID) with $\mathrm{E}_{B-V}$.} 
The panels show EW(NaID) versus the color excess measured towards the stars (left), the narrow lines (middle), and the broad lines (right).
Only sources with detected NaID absorption are shown.
The correlation coefficients and their p-values are presented in the panels.
 }\label{f:naid_vs_dust}
\end{figure*}

In figure \ref{f:naid_vs_dust} we show the correlations of EW(NaID) with the color excess measured towards the stars, the narrow lines, and the broad lines.

\section{Correlations of outflow properties with $L_{\mathrm{AGN}}$ and $L_{\mathrm{SF}}$}\label{a:corr_to_SF_and_AGN}

\begin{figure*}
	\centering
\includegraphics[width=0.7\textwidth]{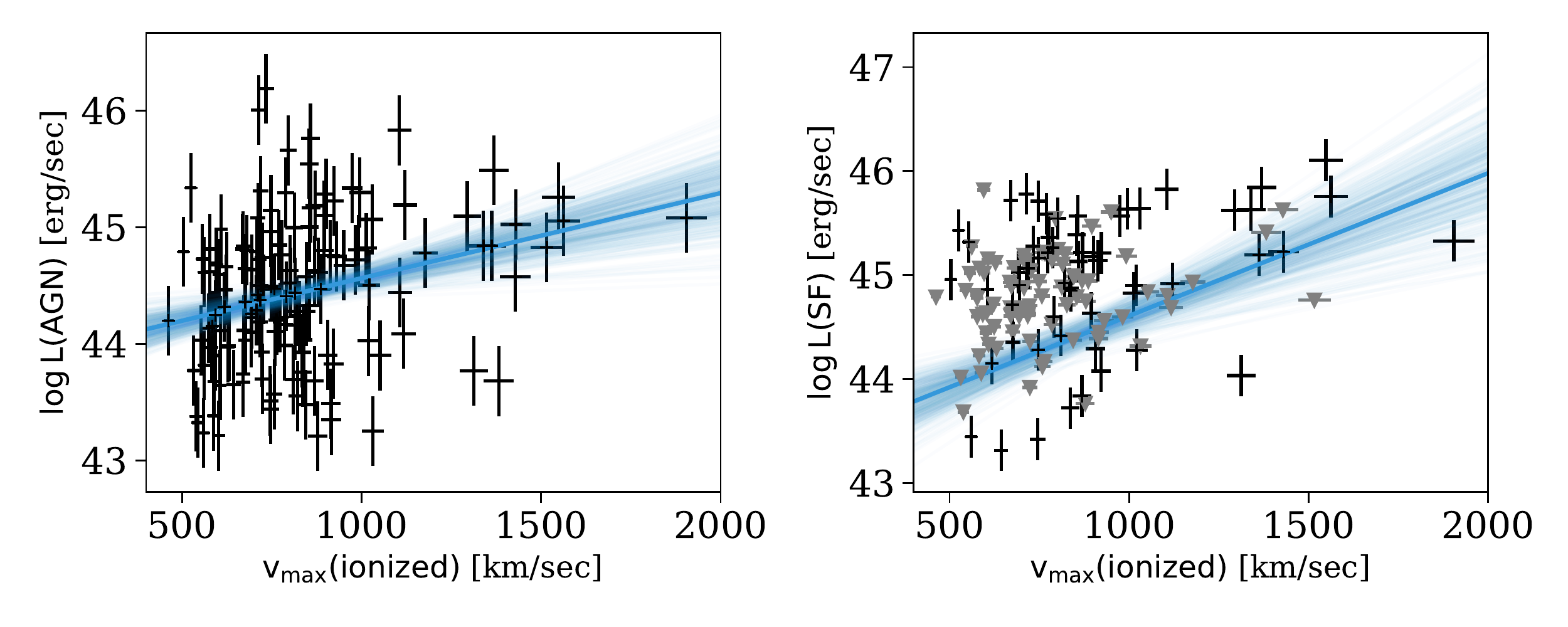}
\includegraphics[width=0.7\textwidth]{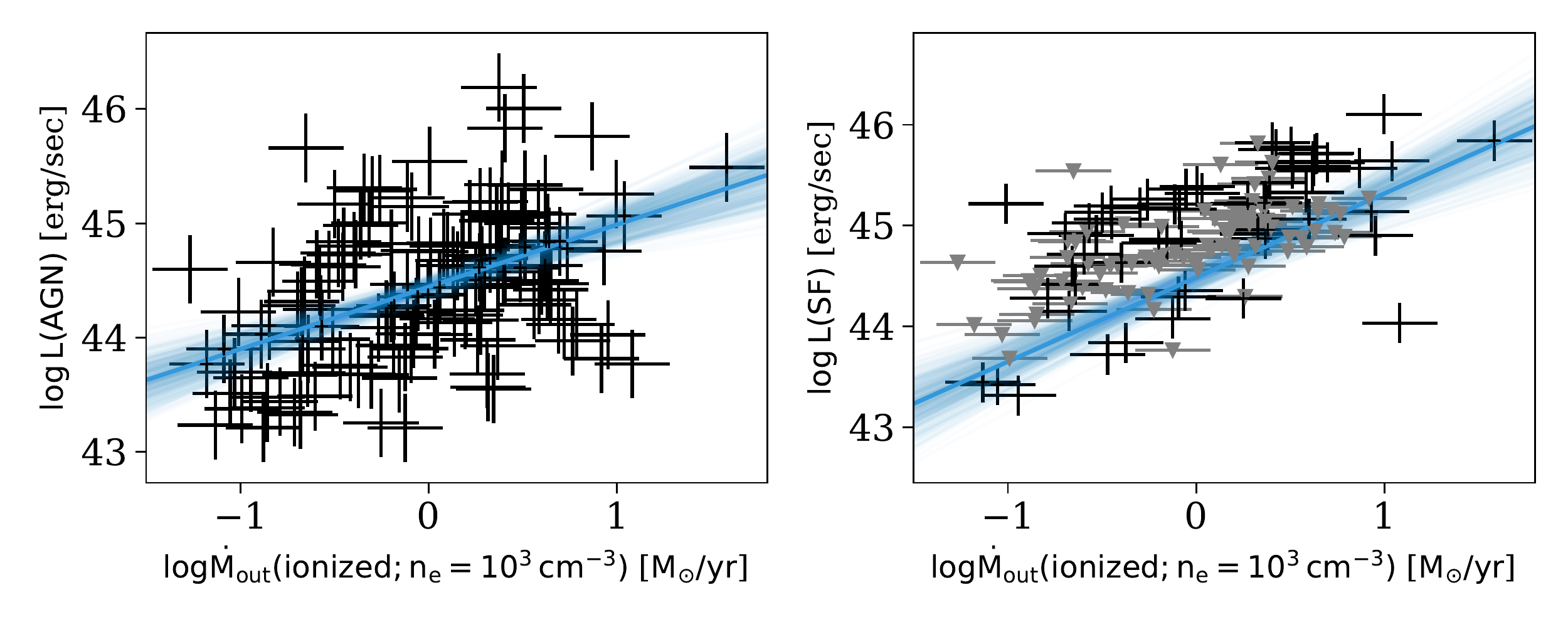}
\includegraphics[width=0.7\textwidth]{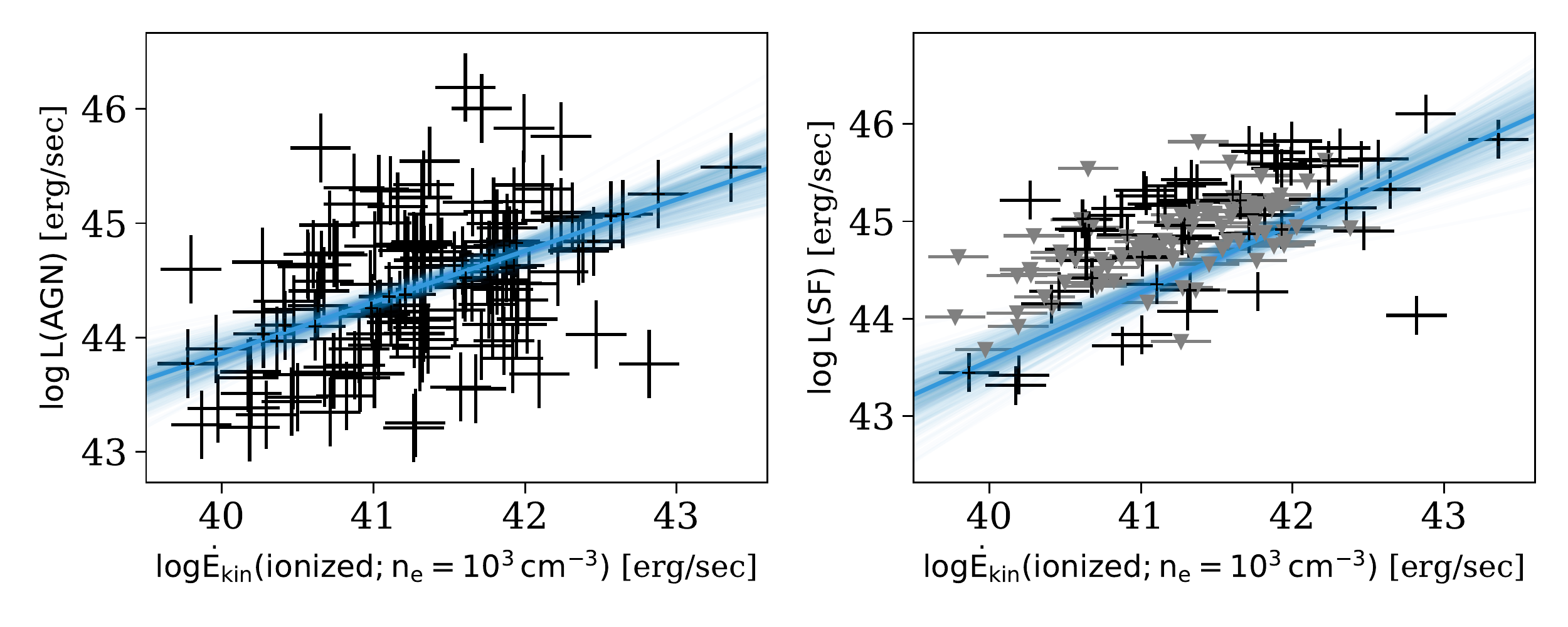}
\caption{\textbf{Correlations of ionized outflow properties with $L_{\mathrm{AGN}}$ and $L_{\mathrm{SF}}$.} 
Each row presents a correlation between an ionized outflow property and $L_{\mathrm{AGN}}$ (left) or $L_{\mathrm{SF}}$ (right). Measured values are marked with black and upper limits with grey. 
The blue lines represent the best-fitting linear relations obtained using {\sc linmix}, and the fainter blue lines represent the uncertainty of the fit. The best-fitting parameters are listed in table \ref{tab:lsf_lagn_corrs}.
 }\label{f:LSF_and_LAGN_versus_ionized_outflow}
\end{figure*}

\begin{figure*}
	\centering
\includegraphics[width=0.7\textwidth]{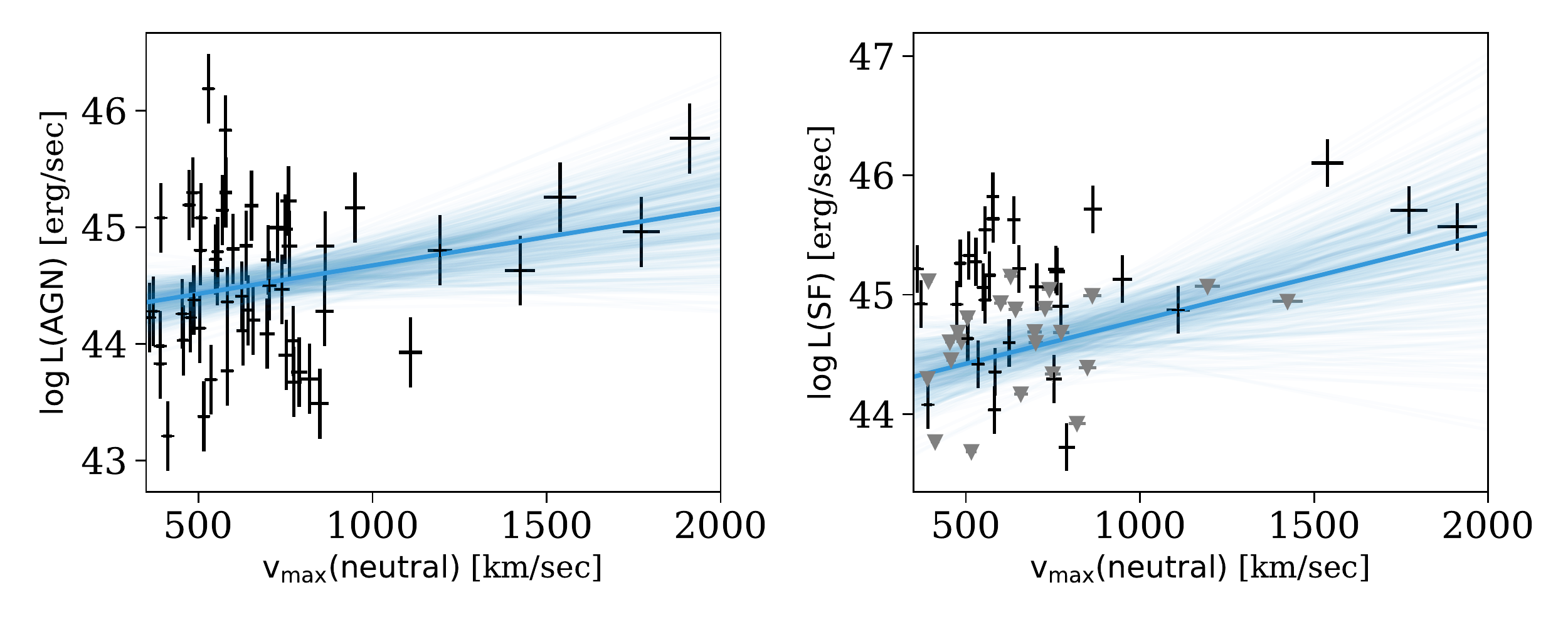}
\includegraphics[width=0.7\textwidth]{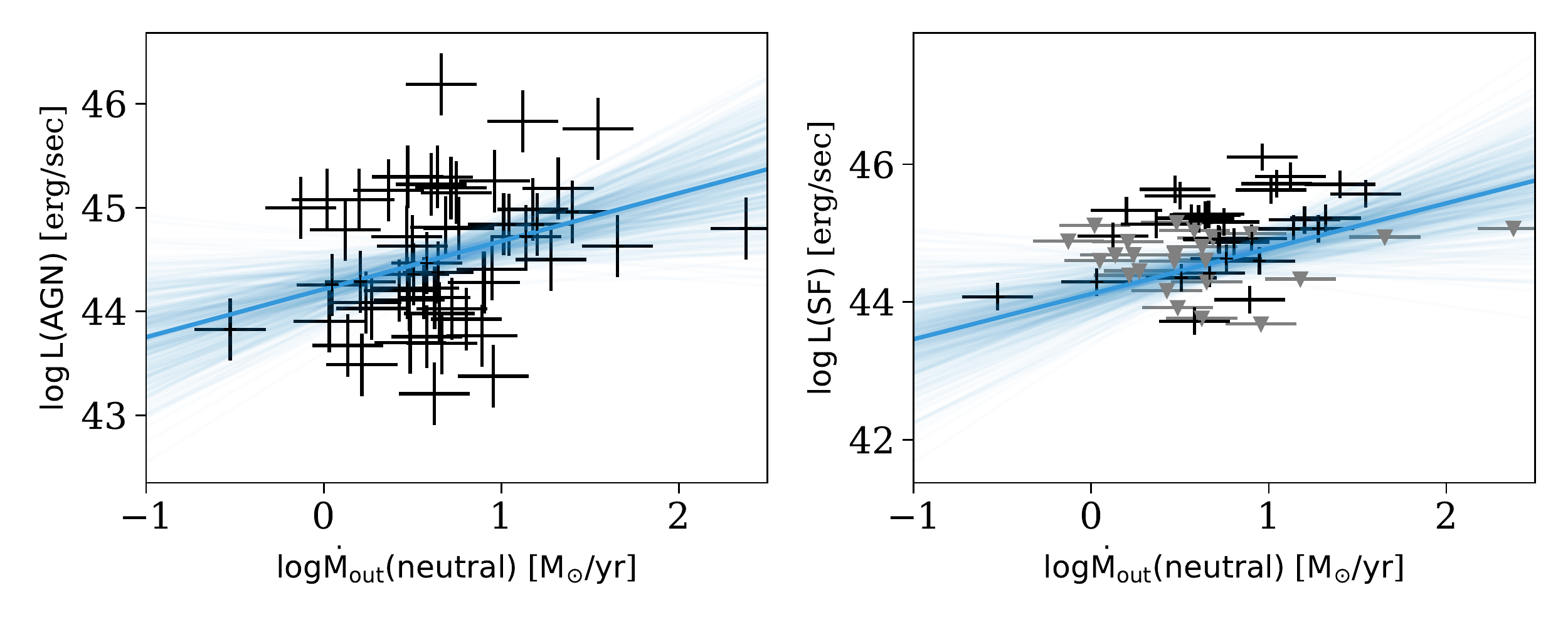}
\includegraphics[width=0.7\textwidth]{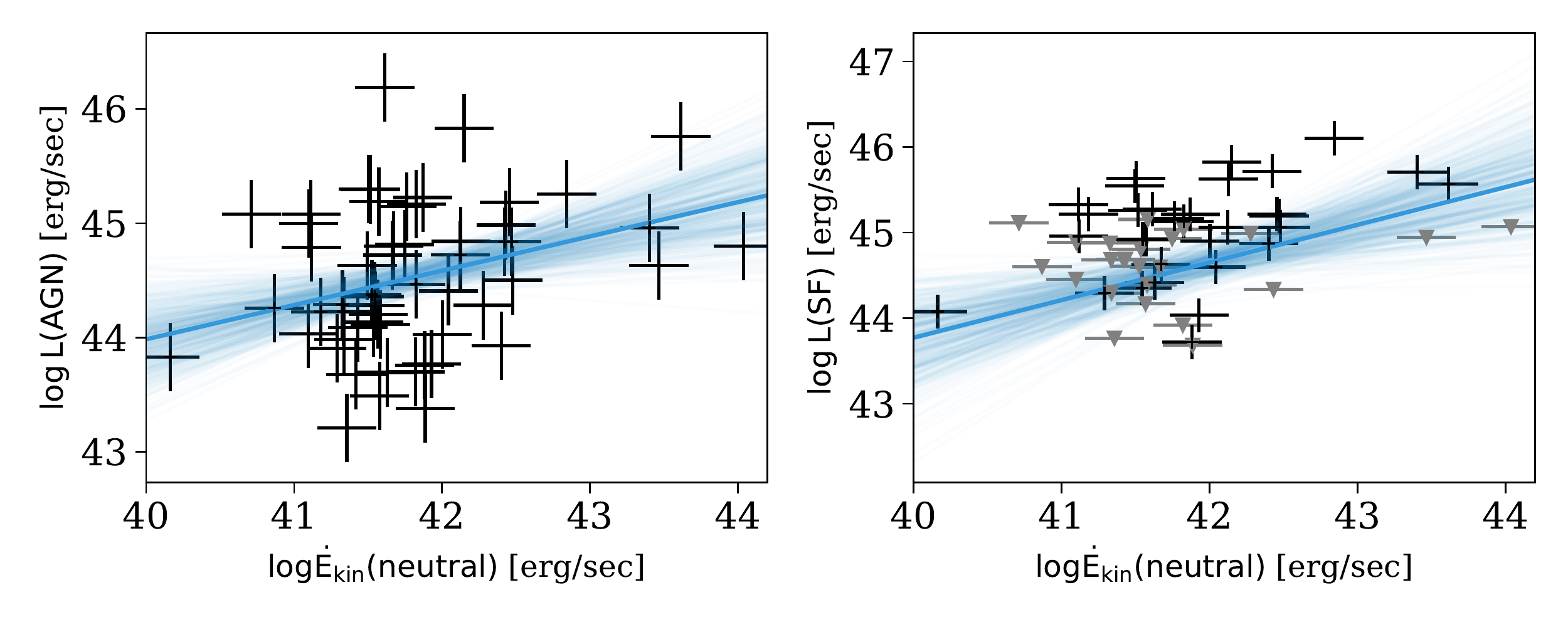}
\caption{\textbf{As figure \ref{f:LSF_and_LAGN_versus_ionized_outflow} but for neutral outflows. } 
 }\label{f:LSF_and_LAGN_versus_neutral_outflow}
\end{figure*}

In table \ref{tab:lsf_lagn_corrs} we list the Kendall's rank correlation coefficients, their associated p-values, and the best-fitting parameters of the linear regression between various outflow properties and $L_{\mathrm{AGN}}$ and $L_{\mathrm{SF}}$. In figures \ref{f:LSF_and_LAGN_versus_ionized_outflow} and \ref{f:LSF_and_LAGN_versus_neutral_outflow} we show the correlations of ionized and neutral outflow properties with $L_{\mathrm{AGN}}$ and $L_{\mathrm{SF}}$.

\begin{table*}
 \caption{\textbf{Correlations between outflow properties and $\mathrm{L_{AGN}}$ and $\mathrm{\mathrm{L_{SF}}}$.} 
For each pair of properties we list the Kendall's rank correlation coefficient $\tau$ and its associated p-value. 
For the significant and marginally significant correlations we also performed linear regression using {\sc linmix} and list the best-fitting parameters, where: $y = A + Bx$. 
 }\label{tab:lsf_lagn_corrs}
 	\centering
\begin{tabular}{|l l| c c | c c |}
\hline
   &   & \multicolumn{2}{c|}{Kendall's rank correlation} & \multicolumn{2}{c|}{{\sc linmix} regression} \\
 x & y & $\tau$ & p-value & A & B \\
\hline
 \multicolumn{6}{|c|}{ionized outflow} \\
\hline
$\mathrm{v_{out}}$ & $\mathrm{\log{} L_{AGN}}$ & 0.21 & 0.00027 & $43.82 \pm 0.18$ & $0.00073 \pm 0.00021$ \\
$\mathrm{v_{out}}$ & $\mathrm{\log{} L_{SF}}$  & 0.16 & 0.0050  & $43.23 \pm 0.29$ & $0.00136 \pm 0.00031$ \\
\hline
$\mathrm{\log{} \dot{M}_{out}}$ & $\mathrm{\log{} L_{AGN}}$ & 0.29 & $1.4 \times 10^{-7}$ & $44.445 \pm 0.047$ & $0.543 \pm 0.094$ \\
$\mathrm{\log{} \dot{M}_{out}}$ & $\mathrm{\log{} L_{SF}}$ & 0.25 & $9.6 \times 10^{-6}$ & $44.480 \pm 0.074$ & $0.83 \pm 0.12$ \\
\hline
$\mathrm{\log{} \dot{E}_{out}}$ & $\mathrm{\log{} L_{AGN}}$ & 0.32 & $1.3 \times 10^{-8}$ & $25.8 \pm 3.0$ & $0.449 \pm 0.072$ \\
$\mathrm{\log{} \dot{E}_{out}}$ & $\mathrm{\log{} L_{SF}}$ & 0.25 & $8.0 \times 10^{-6}$ & $15.5 \pm 4.0$ & $0.701 \pm 0.097$ \\
\hline

 \multicolumn{6}{|c|}{neutral outflow} \\
\hline
$\mathrm{v_{out}}$ & $\mathrm{\log{} L_{AGN}}$ & 0.074 & 0.42 &  &  \\
$\mathrm{v_{out}}$ & $\mathrm{\log{} L_{SF}}$  & 0.079 & 0.38  &  & \\
\hline
$\mathrm{\log{} \dot{M}_{out}}$ & $\mathrm{\log{} L_{AGN}}$ & 0.20 & 0.030 & $44.20 \pm 0.16$ & $0.47 \pm 0.21$ \\
$\mathrm{\log{} \dot{M}_{out}}$ & $\mathrm{\log{} L_{SF}}$ & 0.20 & 0.022 & $44.09 \pm 0.25$ & $0.67 \pm 0.30$ \\
\hline
$\mathrm{\log{} \dot{E}_{out}}$ & $\mathrm{\log{} L_{AGN}}$ & 0.19 & 0.035 & $32.1 \pm 5.5$ & $0.29 \pm 0.13$ \\
$\mathrm{\log{} \dot{E}_{out}}$ & $\mathrm{\log{} L_{SF}}$ & 0.19 & 0.040 & $26.4 \pm 7.4$ & $0.43 \pm 0.17$ \\
\hline

\end{tabular}
\end{table*}

\end{document}